\documentclass[a4paper,11pt]{article}
\pdfoutput=1 % if your are submitting a pdflatex (i.e.~if you have
\usepackage{jheppub} % for details on the use of the package, please see the JHEP-author-manual

\usepackage{booktabs,multirow, quotes}

\usepackage[english]{babel}
\usepackage{amsthm,amsmath,amssymb,graphicx,xcolor,alltt}
\usepackage{framed}

\newcommand{\beq}{\begin{equation}}
\newcommand{\eeq}{\end{equation}}
\newcommand{\bea}{\begin{eqnarray}}
\newcommand{\ea}{\end{eqnarray}}

\theoremstyle{remark}

\theoremstyle{theorem}

\makeatletter
\def\@fpheader{\ }
\makeatother

\title{Direct tests of conformal symmetry with Rydberg atoms: assessment and proposals}

\author{Junchen Rong$^1$,}
\author{Slava Rychkov$^2$}

\affiliation{$^1$CPHT, CNRS, \'Ecole Polytechnique, Institut Polytechnique de Paris, Palaiseau, France\\
$^2$Institut des Hautes \'Etudes Scientifiques, 91440 Bures-sur-Yvette, France}

\emailAdd{junchen.rong@polytechnique.edu}
\emailAdd{slava@ihes.fr}

\abstract{
We propose a direct test of conformal symmetry using Rydberg-atom arrays. By arranging the atoms on a circle and measuring equal-time two-point correlation functions, the characteristic functional form predicted by conformal field theory can be tested directly. We estimate the number of experimental snapshots required to resolve this behavior with sufficient statistical precision. We further optimize the time-dependent ramp profile of the Rabi frequency~$\Omega$ and detuning~$\Delta$ used to prepare the target critical state adiabatically. The optimized profile achieves higher fidelity over much shorter evolution times than those employed in previous experiments. Taken together, these results suggest that a direct experimental test of conformal symmetry is feasible with current Rydberg-atom technology.}

\begin{document} 
	
	\maketitle
	
	\flushbottom

\flushbottom

\section{Introduction}

Conformal invariance as a symmetry of critical fluctuations was first
conjectured by Polyakov in 1970 {\cite{Polyakov:1970xd}}. This idea was one of
the main motivations for the development of the conformal field theory (CFT),
by Polyakov and others {\cite{Polyakov:1974gs,Belavin:1984vu}}. By now it is
widely believed that a great majority of critical points---both thermodynamic
and quantum---possess an emergent conformal invariance, so that their
long-distance behavior is described by a CFT
{\cite{Cardy1987,Cardy:1996xt,tsvelik,giamarchi2003quantum,sachdev,fradkin}}.\footnote{Well-understood
	reasons {\cite{Polchinski:1987dy,Nakayama:2013is}} do allow rare exceptions
	--- critical points with only scale symmetry.} CFTs are tightly constrained
and can be solved, in low dimensions often exactly
{\cite{DiFrancesco:1997nk}}, while in higher dimensions approximately, using
the numerical conformal bootstrap {\cite{Poland:2016chs,Poland:2018epd,Poland:2022qrs,Rychkov:2023wsd}}. Since the scaling
dimensions of CFT local operators are related to the critical exponents, which
can be experimentally measured, this provides many
\text{{\bfseries{indirect}}} tests of conformality,\footnote{We will use
	`conformal symmetry', `conformal invariance' and `conformality'
	interchangeably.} as reviewed e.g.~in {\cite{Henkel1999,Rychkov:2025zks}}. In
comparison, until recently little theoretical effort was devoted to designing
\text{{\bfseries{direct}}} tests {\cite{Pokrovskii1973,Gompper1985}}, and such
experiments were lacking {\cite{Rychkov:2025zks}}. For
\text{{\bfseries{thermodynamic}}} phase transitions, such tests are
challenging due to limitations of imaging with X-ray or neutron probes, although see
{\cite{Podo:2026hfh}} for a recent proposal to use grazing scattering of
X-rays on critical binary alloys.

In the meantime, cold atom platforms {\cite{Browaeys:2020kzz,Morgado:2020jfo}}
revolutionized the experimental study of \text{{\bfseries{quantum}}} phase
transitions
{\cite{Bernien:2017ubn,Scholl:2020hzx,Ebadi:2020ldi,Semeghini:2021wls}}. Many
quantum Hamiltonians of interest can thus be engineered and their quantum
critical points can be studied
{\cite{Keesling:2018ish,Fang:2024uyf,Sun:2026aqf}}. These experiments can in
principle measure the state of every atom, providing a snapshot view of the
many-body wavefunction. Averaging over many snapshots allows to measure e.g.
the equal-time correlation functions
{\cite{Fang:2024uyf,Emperauger:2025raf}}---a classic way to probe conformality
{\cite{Polyakov:1970xd}}.

The time is therefore ripe for a general discussion of existing and future
tests of conformality in cold atom experiments, which will be our purpose
here. Our paper has two main parts. In Section \ref{sec:gen}, we
describe our view of which tests should count as indirect and which ones are
direct. Indirect tests are measurements of critical exponents, while direct
tests need to verify that correlation functions satisfy a conformal Ward
identity. There are also more subtle tests of the so-called state-operator correspondence (SOC), which we also count as direct.
We illustrate these points by examples of existing experiments. In fact, one recent experiment \cite{Sun:2026aqf} performed a direct test of conformality via testing SOC. On the other hand, measurements of conformal correlators have so far been difficult due to decoherence effects producing a finite correlation length smaller than the system size.

This leads us naturally to Section \ref{sec:prop}, where we revisit the simplest setup to test
conformality, through a measurements of equal-time two-point (2pt) correlation
function with periodic boundary conditions in 1D. We present results of
our numerical simulations for experiments with 1D chains of Rydberg atoms. Our main proposal is how to optimize the ramp profiles for the approximate ground state preparation. We show that optimization allows to get ramps with much shorter evolution times, while the fidelity stays the same or higher than in the previous experiments \cite{Fang:2024uyf}.  The use of such optimized ramps should reduce decoherence effects. 

Our conclusions (Section \ref{sec:concl}) are twofold. First, we hope to have convinced the readers that direct tests of conformality via measuring correlators is a worthwhile enterprise. Second, such tests may well be within reach, since simply repeating the existing experiments with the optimized ramps may already yield them.

{\bf Notation:} As is
common in the condensed matter and experimental literature, we use D for the spatial dimensionality, while d stands for the full spacetime dimension, 

\section{Tests of conformal invariance}\label{sec:gen}

\subsection{General discussion and the main direct test}

We find it convenient to divide possible experimental tests of conformality
into two groups---indirect and direct.

By means of analogy, suppose we have a 3D object and we wish to check its
rotational symmetry, i.e.~if is it a round ball. A direct way would be to
stare at the object from different directions. An indirect way would be to
check the circumference to diameter ratio---does it equal $\pi$? Our point here is
that symmetry is fundamentally about \text{{\bfseries{shapes}}} of objects. In
a quantum field theory the most basic object is a correlator. Hence, a direct
test of conformality should test constraints which it imposes on the
functional form of correlators. On the other hand, an indirect test can be a
measurement of a number predicted using techniques of CFT. Indeed, specific CFTs can
often be solved exactly, or numerically to very high accuracy. For example, an
infinite series of 2d CFTs called minimal models was solved exactly,
determining the local operator scaling dimensions and other properties
{\cite{Belavin:1984vu,DiFrancesco:1997nk}}. The main critical exponents
characterizing phase transitions can be expressed in terms of these scaling
dimensions {\cite{cardy1996scaling}}. Then, any measurement of a critical experiment
in agreement with a CFT prediction is an indirect test of conformality. Many
such indirect tests can be extracted from the literature
{\cite{Henkel1999,Rychkov:2025zks}}, even though the original experimental
papers may not have emphasized the connection with conformality.

Talking about CFT correlators, we should specify the geometry, including the
signature: Euclidean or Lorentzian. The Euclidean signature is relevant for the
description of thermodynamic phase transitions, and the Lorentzian signature for
the quantum ones. As a representative example, consider the $\mathcal{M} (3,
4)$ minimal model, also called the 2d Ising CFT, since it describes the
universality class of the Ising phase transition. This theory has local
primary operators $\sigma, \epsilon$ of scaling dimensions $\Delta_{\sigma} =
1 / 8$, $\Delta_{\epsilon} = 1$. In infinite flat 2d space, the simplest
nontrivial correlators are two-point (2pt) and three-point (3pt) functions of
these primary operators:
\begin{align}
	\langle \mathcal{O}(x_1)\mathcal{O}(x_2) \rangle_{\mathbb{R}^2} & =
	\frac{1}{x_{12}^{2 \Delta_{\mathcal{O}}}}  \qquad (\mathcal{O}= \sigma,
	\epsilon) \hspace{0.17em},  \label{eq:two-point}\\
	\langle \sigma (x_1) \sigma (x_2) \epsilon (x_3) \rangle_{\mathbb{R}^2} & =
	\frac{C_{\sigma \sigma \epsilon}}{x_{12}^{2 \Delta_{\sigma} -
			\Delta_{\epsilon}} x_{31}^{\Delta_{\epsilon}} x_{32}^{\Delta_{\epsilon}}}
	\hspace{0.17em}, \quad C_{\sigma \sigma \epsilon} = 1 / 2. 
	\label{eq:three-point}
\end{align}
These formulas are valid in both signatures, interpreting $x^2_{ij}$ as either
Euclidean or Lorentzian distance squared between $x_i$ and $x_j$.\footnote{In
	the latter case, the distances are computed with infinitesimal imaginary parts
	for time differences, depending on the operator ordering, giving phases for
	light-like point separations. See e.g. {\cite{Hofman:2008ar}}.}

Which direct or indirect tests of conformality can be done, in principle, using
these formulas?

A direct test would be to measure the 3pt function \eqref{eq:three-point} and
see the given specific functional form in terms of just three parameters
$\Delta_{\sigma}, \Delta_{\epsilon}, C_{\sigma \sigma \epsilon}$. Indeed this
dependence was one of the earliest nontrivial predictions of conformality {\cite{Polyakov:1970xd}}, obtained from the covariance under conformal
maps, while scale symmetry alone would allow an infinite number of
possibilities {\cite{Rychkov:2016iqz}}. Measurements of infinite-volume 3pt
functions are complicated, and they were never attempted.\footnote{We thank
	Thomas Chalopin and David Cl{\'e}ment for a discussion of the possibility of
	realizing such a measurement in their experimental cold-atom platform
	{\cite{Allemand:2025pdq}}.} 

\begin{figure}[h]\centering
	\begin{center}
		\resizebox{200pt}{!}{\includegraphics{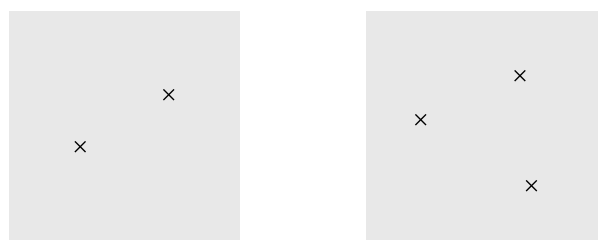}}
	\end{center}
	\caption{\label{fig:flat}  Direct tests in infinite space: 2pt functions (left) test but scale
		symmetry, while 3pt functions (right) do test conformality but are hard to
		measure.}
\end{figure}

On the other hand, to measure the 2pt functions \eqref{eq:two-point} and see a
powerlaw would not be a direct test of conformality, but rather of the more
mundane scale symmetry.\footnote{2pt functions of primaries of unequal
	dimensions vanish by conformality {\cite{Polyakov:1970xd}}, but it's unclear
	how to test this experimentally. See {\cite{Podo:2026hfh}} for a critique of
	an early proposal by Pokrovsky
	{\cite{Pokrovskii1973,PatashinskiPokrovsky1979}}.} See Fig.~\ref{fig:flat}.

Finally, one could use \eqref{eq:two-point} or \eqref{eq:three-point} to
measure $\Delta_{\sigma}$, $\Delta_{\epsilon}$, or $C_{\sigma \sigma
	\epsilon}$ in agreement with their exact CFT values. This would be an indirect
test.

Let us now move to the finite volume case. Rydberg atom experiments are
typically done on groups of 20-200 atoms grouped in a 1D or 2D array. So for
us the finite volume case will be the most relevant. We will be comparing with
2d CFT, so we consider the 1D setup. The most convenient case is that of
periodic boundary conditions, when a chain of Rydberg atoms is arranged
periodically on a circle. We will also comment on the open boundary condition
case, when Rydberg atoms are arranged on an interval.

\begin{figure}[h]\centering
	\begin{center}
		\resizebox{100pt}{!}{\includegraphics{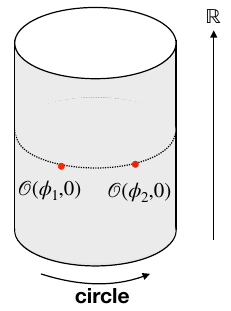}}
	\end{center}
	\caption{\label{fig:cyl}A direct test of conformality in finite volume, using the equal
		time CFT 2pt function on the cylinder $S^1 \times \mathbb{R}$.}
\end{figure}

Let us assume that the chain of Rydberg atoms finds itself in the ground
state, with interactions tuned to be at a quantum critical point. How to get
such a state efficiently via adiabatic evolution will be discussed below. The
corresponding CFT geometry is the surface of a cylinder, where time runs
vertically, and constant-time circular sections represent space where the
Rydberg atoms are positioned, Fig.~\ref{fig:cyl}.

Now, it turns out that such a 1D finite-volume setup carries a theoretical
advantage for direct tests of conformality. We have seen that 2pt functions in
infinite volume would not test conformality but scale symmetry, and one would
have to go to 3pt functions in infinite volume. But in finite circular volume
already 2pt functions test conformality! Indeed, while the scale symmetry is
broken by finite volume, conformality survives. In fact, the conformal group for
the 2d cylinder (or on any geometry which is, like the cylinder, equivalent to
the plane by a Weyl transformation) is as large as the conformal group of the
plane. This leads to constraints on correlators, which are nontrivial already
for 2pt functions. The 2pt function of a primary operator on the cylinder of
circumference $2 \pi$ is given by {\cite{Luscher:1974ez}}:
\begin{equation}
	\label{eq:2ptcyl} \langle \mathcal{O}(\phi_1, \tau_1)\mathcal{O}(\phi_2,
	\tau_2) \rangle_{\text{cyl}} = \frac{1}{|2 \cosh
		(\tau_1 - \tau_2) - 2 \cos (\phi_1 - \phi_2) |^{\Delta_{\mathcal{O}}}}
	\hspace{0.17em},
\end{equation}
where $\phi_i \in [0, 2 \pi]$, $\tau_i \in \mathbb{R}$. We wrote this formula
for the Euclidean cylinder, while for the Lorentzian cylinder one has to do
the Wick rotation.

The salient features of Eq. \eqref{eq:2ptcyl} are: time-translation
invariance, and exponential decay at large Euclidean time separation $\tau_1 -
\tau_2$. To find this formula, the easiest is to use the above-mentioned Weyl
equivalence of the cylinder and the plane, realized by the exponential map.
Thus \eqref{eq:2ptcyl} can be easily obtained from the infinite-plane result
\eqref{eq:two-point} {\cite{DiFrancesco:1997nk}}. However, one
could also derive Eq. \eqref{eq:2ptcyl} without using the
Weyl-equivalence, by staying on the cylinder and using the conformal group of
the cylinder.

For the Hamiltonian dynamics in real time, like in the Rydberg atom
experiments, one needs to Wick-rotate \eqref{eq:2ptcyl} to the Lorentzian
signature. Then, the 2pt function exhibits oscillating behavior in the time
direction {\cite{Luscher:1974ez}}. We don't show the full formula here,
because below we will focus on the equal time case, in which case the 2pt
function reduces to
\begin{equation}
	\label{eq:2ptcylequal} \langle \mathcal{O}(\phi_1, 0)\mathcal{O}(\phi_2, 0)
	\rangle_{\text{cyl}} = \frac{1}{|2 \sin \frac{\phi_1 -
			\phi_2}{2} |^{2 \Delta_{\mathcal{O}}}},
\end{equation}
for both the Euclidean and the Lorentzian signature.

We see that the functional form of the finite-volume correlator
\eqref{eq:2ptcylequal} is fixed at all distances, including distances of the
order of the size of the spatial manifold, in terms of just one parameter
$\Delta_{\mathcal{O}}$. We would like to stress here
that this, well-known, fact is a direct consequence of conformality. Therefore, to see this
functional form experimentally would be a direct test of conformality.

\subsection{Other direct tests}

Below we will mostly focus on the possibility of a direct test of Eq.
\eqref{eq:2ptcylequal}. But in this section we will briefly discuss two other
possible direct tests.

\begin{figure}[h]\centering
	\begin{center}
		\resizebox{100pt}{!}{\includegraphics{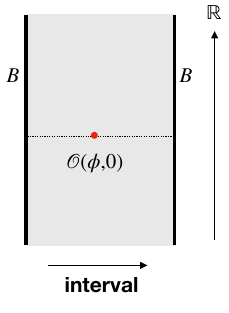}}
	\end{center}
	\caption{\label{fig:strip}A direct test of conformality in finite volume, using the CFT 1pt
		function on the strip $[0, \pi] \times \mathbb{R}$.}
\end{figure}

\text{{\bfseries{I.}}} Instead of periodic boundary conditions (circle), one
may choose to position Rydberg atoms on an interval. Then, the relevant CFT
geometry is the strip, Fig.~\ref{fig:strip}. There is some freedom for the
boundary conditions on the two sides of the strip, playing with interactions
of Rydberg atoms close to the endpoints. At the critical point, the effect of
boundary conditions is described by conformal boundary conditions $B_1$ and
$B_2$ on the two boundaries. In this strip setup, already 1pt functions can
test conformality directly. In the simplest case with equal boundary
conditions, $B_1 = B_2$, the 1pt function of a primary operator on the strip
of width $\pi$ takes the form:
\begin{equation}
	\label{eq:1ptstrip} \langle \mathcal{O}(\phi, 0)
	\rangle_{\text{strip}} = \frac{A_{\mathcal{O}}}{(2 \sin
		\phi)^{\Delta_{\mathcal{O}}}} \hspace{0.17em}, \quad 0 < \phi < \pi
	\hspace{0.17em} \quad (B_1 = B_2) \hspace{0.17em} .
\end{equation}
This is usually shown by conformal-transforming the upper-half-plane 1pt
function $\langle \mathcal{O}(z) \rangle_{\text{UHP}} =
A_{\mathcal{O}} / (2 \hspace{0.17em} \text{Im} z)^{\Delta_{\mathcal{O}}}$
{\cite{Cardy:1984bb}} to the strip {\cite{Calabrese:2006rx}}. But this can
also be in principle derived by staying on the strip.

For a nontrivial test, $A_{\mathcal{O}}$ in \eqref{eq:1ptstrip} should not be
zero, which may happen due to global symmetry reasons. For example, for
$\mathcal{O}= \sigma$ in the 2d Ising CFT, the boundary condition should break
global $\mathbb{Z}_2$ for $A_{\mathcal{O}}$ to be nonzero, hence it should be
either ``spin-up'' or ``spin-down'' boundary condition $| + \rangle$, $| -
\rangle$ and not the ``free'' boundary condition $|f \rangle$.

What is easier: testing 1pt function \eqref{eq:1ptstrip} or testing 2pt
function \eqref{eq:2ptcylequal}? Our investigations suggest that testing
\eqref{eq:2ptcylequal} is easier. A significant advantage of
\eqref{eq:2ptcylequal} is translation invariance. It helps in the measurement,
since averaging over translations improves statistics. It also helps during
the adiabatic evolution when preparing the ground state, since the whole
adiabatic evolution takes place in the translationally invariant subsector of
the Hilbert space. Eq. \eqref{eq:1ptstrip} has no comparable advantages.
Furthermore, in our numerical experiments we saw that \eqref{eq:1ptstrip} is affected by
large corrections to scaling near the boundaries.\footnote{One can phrase this
	by saying that there is an ambiguity where the ``true'' CFT boundary is
	located compared to the Rydberg chain boundary.} For these reasons, below we
will focus on the possible tests of \eqref{eq:2ptcylequal} and we will not
report any results for \eqref{eq:1ptstrip}.

\text{{\bfseries{II.}}} Another possible test of conformality would be to test
the operator-state correspondence (OSC), which can be tested both on the
cylinder and on the strip.

Consider the list $\{ \mathcal{O}_n \}_{n = 0}^{\infty}$ of all local
operators of a CFT, both primaries and their descendants. The OSC on the
cylinder says that \cite{DiFrancesco:1997nk}:
\begin{itemize}
\item
	such operators are in one-to-one correspondence with the energy eigenstates
	$| n \rangle$ on the circle, with the identity operator $\mathcal{O}_0$
	corresponding to the ground state $| 0 \rangle$;
	
	\item scaling dimensions $\Delta_{\mathcal{O}_n}$ and the gaps $E_n - E_0$
	above the ground state are proportional to each other.
\end{itemize}
The proportionality factor involves the ``speed of light'' of the emergent
Lorentz invariance, as well as the physical length of space. To eliminate
these factors, it is convenient to study the energy gap ratios. Then the OSC
predicts that
\begin{equation}
	\label{eq:ratio-rel} (E_n - E_0) / (E_m - E_0) = \Delta_{\mathcal{O}_n} /
	\Delta_{\mathcal{O}_m} \hspace{0.17em} \quad \text{(cylinder)} .
\end{equation}
There is also a variant of OSC for the strip with boundary conditions $B_1$
and $B_2$. There, there is correspondence between boundary (or
boundary-changing if $B_1 \ne B_2$) CFT operators $\hat{\mathcal{O}}_n$ on the
one hand, and states on the interval on the other hand. There is also a
relation analogous to \eqref{eq:ratio-rel} with $\Delta_{\mathcal{O}_n} \to
\Delta_{\hat{\mathcal{O}}_n}$ \cite{DiFrancesco:1997nk}. 

\begin{figure}[h]\centering
	\begin{center}
		\resizebox{200pt}{!}{\includegraphics{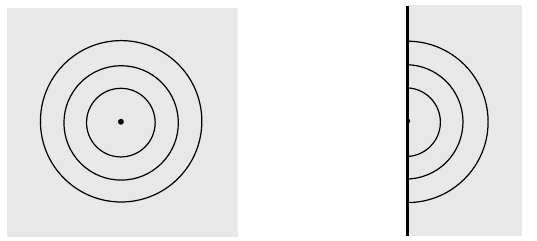}} 
	\end{center}
	\caption{\label{fig:OSC}Justification of OSC. Cylinder, resp. strip, can be conformally
		mapped onto the plane (left), resp. half-plane (right), with constant time
		sections mapped to constant radius (semi)circles.}
\end{figure}

The OSC is a fundamental property of CFT. Its origin can be traced to the
above-mentioned fact that the cylinder is Weyl-equivalent to the plane,
while the strip to the half-plane (see Fig.~\ref{fig:OSC}). Therefore, a test
of relation \eqref{eq:ratio-rel} would also qualify as a direct test of
conformality.\footnote{We thank the authors of {\cite{Wang:2026prw}} for a
	discussion.}

In conclusion, we note that the three above-mentioned 1D finite-volume tests
of conformal invariance---testing \eqref{eq:2ptcylequal}, \eqref{eq:1ptstrip}
or \eqref{eq:ratio-rel}---have no easy counterparts for 2D, where finite
volume typically breaks conformality. The only exception is a two-sphere
spatial manifold. In cold atom experiments, distributing Rydberg atoms on a
two-sphere will break rotational symmetry to a discrete subgroup, which makes
comparison to CFT subtle although not impossible
{\cite{Lao:2023zis,Wu:2026ayb}}. Recently, fuzzy sphere 2D quantum criticality
{\cite{Zhu:2022gjc}}, with electrons moving on the two-sphere with a magnetic
flux, was proposed as a way to preserve rotational invariance. Experimental
realization of this setup is currently lacking.

\subsection{Existing experimental results: Rydberg atoms and related
	platforms}

In this section we will mention several already performed experimental
measurements of quantum criticality with Rydberg atoms and related
experimental platforms, which we find most interesting and/or relevant to the discussion. We will see that one
direct test and several indirect ones can be extracted from these
measurements. We do not discuss here many indirect experimental tests in
thermodynamic phase transitions, which were already reviewed in
{\cite{Henkel1999,Rychkov:2025zks}}.

We start by discussing measurements of critical 2pt functions
{\cite{Fang:2024uyf,Emperauger:2025raf}}. Ref.~{\cite{Fang:2024uyf}} studied
the 2d Ising universality class with $N = 24$ Rydberg atoms in a circle, while
Ref.~{\cite{Emperauger:2025raf}} focused on the XY universality class in the
same geometry. Both works saw a short-distance powerlaw, followed by
exponential decay with correlation length $\xi_{\rm d} = 13.2^{+ 5.7}_{- 3.6}$
{\cite{Fang:2024uyf}} or $\xi_{\rm d} = 15 (4)$ {\cite{Emperauger:2025raf}}. Notably
this correlation length was attributed to decoherence effects and other
imperfections, and not to the lack of adiabaticity. This will play a role in
our proposal in Section \ref{sec:prop}. The important point for now is that, due to this correlation length,
Refs. {\cite{Fang:2024uyf,Emperauger:2025raf}} cannot be said to have
performed a direct test of conformality via Eq.~\eqref{eq:2ptcylequal}. Still,
from the short-distance powerlaw they could extract scaling dimensions and compare
with CFT. Ref.~{\cite{Fang:2024uyf}} saw $\Delta_{\sigma} = 0.127 (37)$ in
agreement with the 2d Ising CFT $1 / 8$. This counts as an indirect test of
conformality. Ref.~{\cite{Emperauger:2025raf}} measured $\Delta \approx 0.13$,
the vertex operator dimension in the compact scalar boson CFT at the Luttinger
parameter $K = 1.6 (4)$. Since $K$ is a modulus (unconstrained parameter), we
do not interpret this result as an indirect test of conformality.

We next discuss Ref.~{\cite{Sun:2026aqf}}, who worked on the interval geometry
(a chain of $N = 19$ Rydberg atoms), and performed energy gap measurements,
using time-periodic modulation. Thus they were able to test the OSC relation
\eqref{eq:ratio-rel} for the strip. For the critical Ising model with spin-up
boundary conditions on both sides of the strip, they measured low energy gaps
up to level $\sim 7$, showing that they are are equally spaced, in agreement
with the free Majorana theory description of the underlying CFT. They next
studied the tricritical Ising (TCI) model with several boundary conditions.
E.g.~for TCI with the free boundary conditions they measured $(E_2 - E_0) /
(E_1 - E_0) = 1.45 (25)$ in agreement with the CFT prediction $4 / 3$. As the
TCI boundary spectrum does not have a free theory description, this is the
first nontrivial direct test of conformality known to us.

The discussed measurements of
Ref.~{\cite{Fang:2024uyf,Emperauger:2025raf,Sun:2026aqf}} were done with the
critical ground state, prepared via adiabatic evolution, which we will discuss
below. There exists another setup: to drive a 1D
{\cite{Keesling:2018ish,Zhang:2025xkp}} or 2D
{\cite{Ebadi:2020ldi,Manovitz:2024hif}} array of Rydberg atoms through the
critical point non-adiabatically, and extract critical exponents from the
Kibble-Zurek (KZ) scaling {\cite{delCampo:2013nla}}. As any exponent
measurement, these count as indirect tests of conformality, including its
prediction $z = 1$ for the dynamical critical exponent. We do not know if one
can get a direct test of conformality in a KZ setup.

The KZ scaling for the 1D transverse field Ising model driven through the
quantum critical point was also studied using a D-Wave quantum annealer
{\cite{King:2022phl}}. However we are not aware of correlator measurements
using this experimental platform.

We also mention closely related digital quantum computer studies
{\cite{Anand:2022cdi,Dborin:2022zdd,Haghshenas:2023bje}}. They first use a
classical computer to find a holographic quantum circuit describing the
critical state of interest, e.g.~the 2d Ising criticality. The circuit is then
evaluated on a quantum computer, measuring e.g.~the 2pt function. Because of
their hybrid character, these works are not fully experimental tests of
conformality in a quantum phase transition. However they do demonstrate
growing power of quantum computing.

\section{Critical 2pt function on a circle with Rydberg atoms}\label{sec:prop}

In the rest of the paper we will discuss the prospects of measuring the
functional dependence ~\eqref{eq:2ptcylequal} with Rydberg atoms, thus
performing a direct test of conformality. As mentioned above, the critical 2pt
function on the circle was already measured in
{\cite{Fang:2024uyf,Emperauger:2025raf}}. But they did not see Eq.
\eqref{eq:2ptcylequal}. Instead they saw an exponential decay. Below we will
argue that simple improvements of these existing experiments have a good
chance to reduce decoherence sufficiently so that Eq. \eqref{eq:2ptcylequal}
can be verified.

This section is structured as follows. In Section \ref{sec:hamiltonian-and-phase-diagram} we describe the Rydberg spin chain Hamiltonian, its phase diagram, and choose two points on the Ising critical line which will be studied in more detail below. In Section \ref{sec:measuring-critical-2pt-point-function} we describe the main steps of how a critical 2pt function is measured, and the main issues affecting this measurement. These issues are discussed one-by-one in subsequent Sections \ref{sec:2ptDMRG}-\ref{sec:decoherence}. 

\subsection{Hamiltonian and phase diagram}\label{sec:hamiltonian-and-phase-diagram}

We consider the standard setup known to lead to the Ising quantum criticality in
1D chains of Rydberg atoms {\cite[Eq.~(1)]{Browaeys:2020kzz}}. As mentioned we
are interested in the periodic circular arrangements. In this section we
consider the ideal setup where the atoms are exactly equally spaced. The
effects of atoms not exactly equally spaced will be mentioned below.

\begin{figure}[h]\centering
\includegraphics[width=100pt]{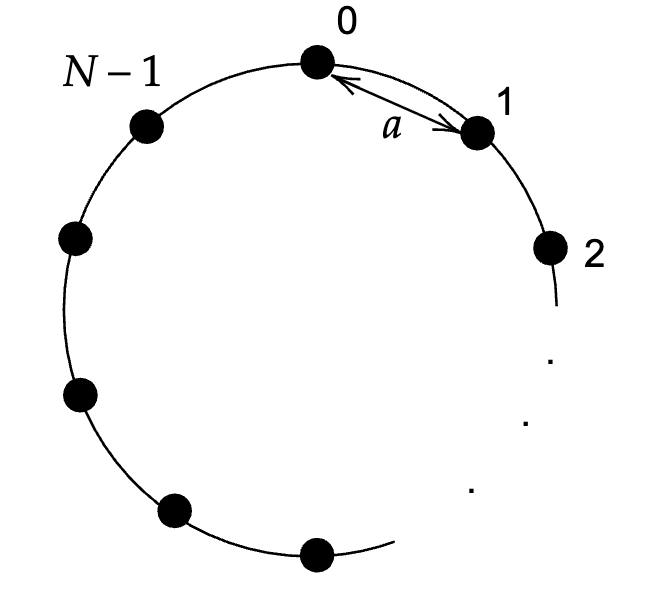}
	\caption{\label{fig:chain}$N$ Rydberg atoms placed on a circle.}
\end{figure}

To write the Hamiltonian, we identify the Hilbert space of the Rydberg chain
with the Hilbert space of a spin-1/2 spin chain, identifying the ground state
$|0\rangle_i$ and the Rydberg state $|1\rangle_i$ of each atom with down and
up spin-$1 / 2$ states $|\!\!\downarrow\rangle_i, |\!\!\uparrow\rangle_i$ of the
corresponding spin. The Hamiltonian for the chain of $N$ atoms (Fig.~\ref{fig:chain}) then becomes
\begin{equation}
	H_{\text{phys}} = \frac{\hbar \Omega}{2}  \sum_{i = 0}^{N - 1} X_i -
	\hbar \Delta \sum_{i = 0}^{N - 1} n_i + \sum_{i < j} V_{ij} n_i n_j, \quad
	V_{ij} = \frac{C_6}{R_{ij}^6} . \label{eq:model}
\end{equation}
Here $\Omega$ is the effective Rabi frequency, $\Delta$ is the detuning, $n_i
= |1\rangle_i \langle 1|_i$ counts Rydberg excitations at site $i$, $X_i$ is
the Pauli-x matrix, $R_{ij} = | \mathbf{r}_i - \mathbf{r}_j |$ is the distance
between the atoms, and $V_{ij}$ is the van der Waals interaction assumed
repulsive, $C_6 > 0$.

Denoting $a$ the distance between neighboring atoms, and \ $U = C_6 /
a^6$, we will work with the dimensionless Hamiltonian $H =
H_{\text{phys}} / U$:
\begin{equation}
	H  = \frac{\Omega'}{2}  \sum_i X_i - \Delta'  \sum_i n_i + \sum_{i <
		j} \frac{1}{d_{ij}^6} n_i n_j,  \label{eq:modelp}
\end{equation}
where $\Omega' = \hbar \Omega / U$ and $\Delta' = \hbar \Delta / U$ are real
dimensionless numbers, and $d_{ij} = R_{ij} / a$ is the dimensionless
distance. 

\begin{figure}[h]\centering
	\begin{center}
		\resizebox{0.55\columnwidth}{!}{\includegraphics{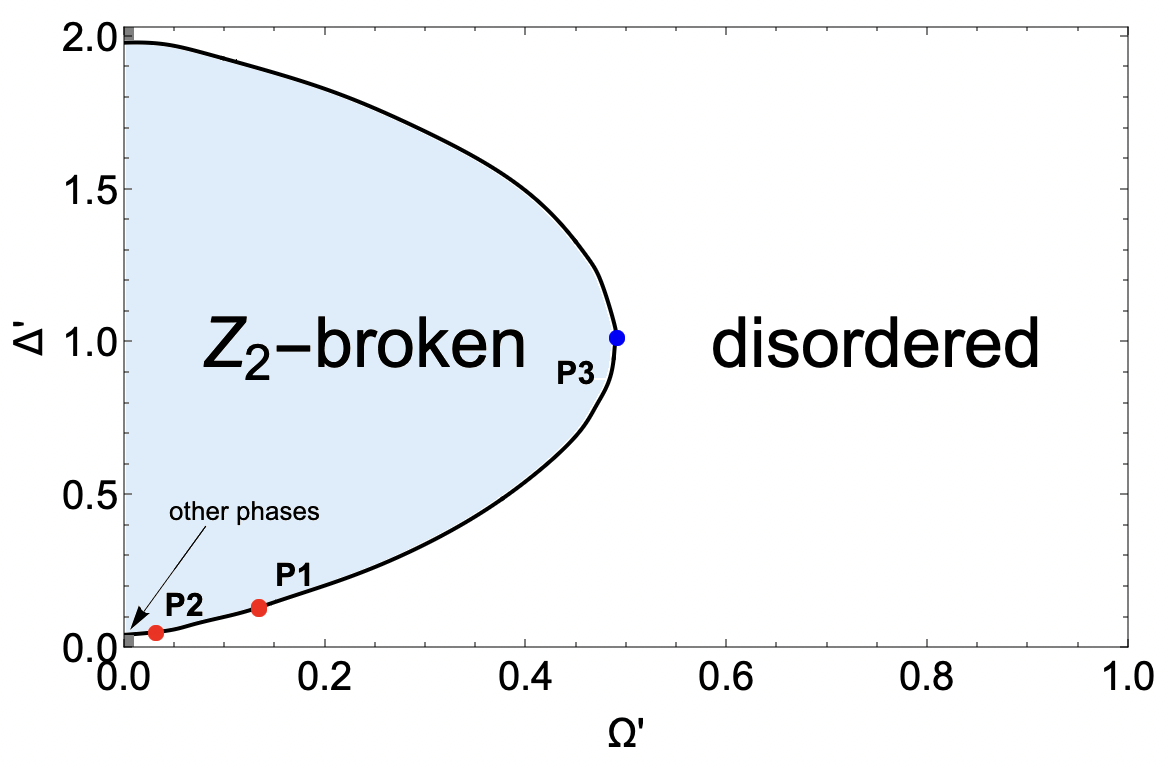}}
	\end{center}
	\caption{\label{phasediagramHBlong}A part of the phase diagram of the model \eqref{eq:model} showing
		the Ising phase transition between the $\mathbb{Z}_2$-broken ``density wave'' phase
		and the disordered phase. We do not show in detail the two tiny gray regions
		which house further phases {\cite{Keesling:2018ish,Rader:2019syq}}. The two
		red points corresponds to the points studied in~{\cite{Fang:2024uyf}} (P1)
		and {\cite{Sun:2026aqf}} (P2), see Table {\color{red}{\ref{tab:params}}}.
		Below we will focus on point P1 as well as the blue point P3 corresponding
		to $\Delta' = \zeta (6)$, $\Omega' \approx 0.488$.}
\end{figure}

Let us describe the phase diagram of this Hamiltonian. It is rich and well
studied {\cite{Keesling:2018ish,Rader:2019syq}}. Most of the parameter space
(Fig.~\ref{phasediagramHBlong}) is occupied by two non-critical phases: 1) the
disordered phase and 2) the density wave phase which breaks spontaneously
$\mathbb{Z}_2$ translation by one lattice spacing. These two phases are
separated by the line of critical points, all belonging to the Ising
universality class. This critical line will be our main interest. The
mentioned features of the phase diagram can be understood rewriting the
Hamiltonian in terms of Pauli matrices via $n_i = (1 + Z_i) / 2$:
\begin{equation}
	H = \frac{\Omega'}{2}  \sum_i X_i + \sum_{i < j} \frac{1}{4 d_{ij}^6}
	Z_i Z_j + \sum_i \left( \sum_j \frac{1}{4 d_{ij}^6} - \frac{\Delta'}{2}
	\right) Z_i +\text{const}
\end{equation}
Then, if $\Omega', \Delta'$ are not too small, we can neglect the
non-nearest-neighbor $Z_i Z_j$ couplings. This gives the antiferromagnetic
Ising model in longitudinal and transverse magnetic fields, which has a very
similar phase diagram {\cite{Ovchinnikov}}. In particular, in the
infinite-volume limit, the longitudinal field vanishes for $\Delta' = \zeta
(6) \approx 1.017$. The phase diagram is symmetric with respect to this line,
as well as with respect to $\Omega' \to - \Omega'$ (so that we only show
$\Omega' > 0$).

The above features of the phase diagram will be sufficient for us, but for
completeness we describe the additional featurers of the phase diagram in the
strong Rydberg blockade regime $\Omega', \Delta' \ll 1$, which requires a
separate treatment. This regime can be studied via the effective
Fendley-Sengupta-Sachdev (FSS) model {\cite{FSS}} or directly from
\eqref{eq:modelp} {\cite{Keesling:2018ish,Rader:2019syq}}. One finds, for
$\Delta' \lesssim 0.05$, $\Omega' \lesssim 0.01$, a hierarchy of crystalline
phases of period 3,4,5,..., as well as non-commensurate ``floating'' phases
separating them.

As mentioned we will focus on the $\mathbb{Z}_2$ critical line and we will not
be interested in the exotic phases. Everywhere on this line, the phase
transition is in the Ising universality class, although the Hamiltonian
$H$ does not generically have onsite $\mathbb{Z}_2$ symmetry. At a
generic point, the $\mathbb{Z}_2$ symmetry of the Ising CFT emerges (or
``emanates'' {\cite{Seiberg:2023cdc}}) from the one-unit lattice translation
{\cite{Slagle:2021ene}}, which permutes the two ground states in the broken
phase. The only exception is $\Delta' = \zeta (6)$ where the Hamiltonian does
have onsite $\mathbb{Z}_2$: $Z_i \to - Z_i$.

Below for definiteness we will focus on two points on the critical line:
point P3 $\Delta' = \zeta (6)$, $\Omega' \approx 0.488$ which has on-site
$\mathbb{Z}_2$, and point P1 from {\cite{Fang:2024uyf}} (see Table
\ref{tab:params}) which lies closer to (but not quite in) the strong blockade
regime. We will see that P3 has some advantages for the experiments. All
simulations for both points will be done with the Hamiltonian
\eqref{eq:modelp}.

\begin{table}[h]
	\centering
	\begin{tabular}{lll}
		\toprule
		point & $\Omega'$ & $\Delta'$\\\midrule
		P1{\cite{Fang:2024uyf}} & 0.133 & 0.137\\
		P2{\cite{Sun:2026aqf}} & 0.031 & 0.052\\
		P3 & $0.488$ & $\zeta (6)$\\
		\bottomrule
	\end{tabular}
	\caption{\label{tab:params}Parameters $\Omega', \Delta'$ of points P1 (\cite{Fang:2024uyf}, see App.~\ref{sec:Yao-details}) and P2
		\cite{Sun:2026aqf} used in previous experimental studies of 2d Ising criticality with Rydberg
		atoms, as well as of point P3 which has
		on-site $\mathbb{Z}_2$. }
\end{table}

\subsection{Measuring critical 2pt function}\label{sec:measuring-critical-2pt-point-function}

We will be interested in the equal-time CFT 2pt function $\langle
\mathcal{O}(\phi_1, 0)\mathcal{O}(\phi_2, 0) \rangle$, see
Eq.~\eqref{eq:2ptcylequal}, where we take $\mathcal{O}= \sigma$, the
lowest-dimension nontrivial primary of the 2d Ising CFT, which is $\mathbb{Z}_2$-odd. Correlation functions
of the $\mathbb{Z}_2$-even primary $\epsilon$ can also be discussed, but since $\epsilon$ has a higher scaling
dimension, they will decay faster with a distance and the powerlaw will be
harder to resolve experimentally.

In a Rydberg atom experiment, this critical 2pt function is measured as
follows {\cite{Fang:2024uyf}}.
\begin{enumerate}
	\item One starts in the ground state of the Rydberg chain at the initial
	point $(\Omega_0', \Delta_0')$ with $\Omega_0' = 0$, $\Delta_0' < 0$. For
	$\Omega' = 0$, the Hamiltonian trivializes and the ground state is easy to
	find: for any $\Delta' < 0$ it is the tensor product state $| 0
	\rangle^{\otimes N}$. This ground state is also easy to prepare
	experimentally.
	
	\item In the $(\Omega', \Delta')$ plane one follows a curve $(\Omega' (t), \Delta' (t))$, $0 \leqslant t
	\leqslant T$, connecting the initial
	point with a final point chosen to lie on the critical line (for us, P1 or
	P3). This curve is chosen so that the quantum mechanical evolution is
	sufficiently adiabatic so that the final state is sufficiently close to the
	ground state. E.g. the curve used in {\cite{Fang:2024uyf}} is shown in
	Fig.~\ref{fig:evolution}.
		
	\item At time $T$, a projective measurement is performed. The result of this
	measurement is the collapsed state of the form $\Psi  = | n_0, \ldots, n_{N-1}
	\rangle$ where $n_i = 0$ or 1 if atom $i$ was found in the ground or the
	Rydberg state.
	
	\item Repeating steps 1,2,3 several times, and averaging over the resulting
	collapsed states, one measures the 2pt function of an appropriately chosen
	microscopic operator $s_i$. This operators must be diagonal in the $n_i$
	basis. The operator is chosen so that, at the quantum critical point, and at
	separation $| i - j | \gg 1$, its 2pt function equals (up to position
	independent rescaling), the CFT 2pt function we are interested in:
	\begin{equation}
		\langle s_i s_j \rangle =\text{const.} \langle \sigma
		(\phi_1, 0) \sigma (\phi_2, 0) \rangle, \quad \phi_i = 2 \pi i / N, \quad
		| i - j | \gg 1. \label{ssCFT}
	\end{equation}
\end{enumerate}
Several issues affect the measurement processes, rendering it more or less
experimentally feasible. They are discussed below:
\begin{enumerate}
	\item (Section \ref{sec:2ptDMRG}) At which minimal separation $| i - j |$
	can we trust Eq. \eqref{ssCFT}? We will see that $| i - j | = 1$ is already
	good enough.
	
	\item (Section \ref{sec:sample}) How many collapsed states do we need so
	that statistical errors in the measurement of $\langle s_i s_j \rangle$ are
	reasonably small, say below 10\%? I.e.~how many times do we have to repeat the
	experiment? We will see that order $10^3$ times is enough.
	
	\item (Section \ref{sec:fidelity}) How can we optimize the evolution curve
	$(\Omega' (t), \Delta' (t))$ so that, for fixed evolution time $T$, the
	final state is as close as possible to the true ground state? What are the
	achievable fidelities for experimentally reasonable evolution times? We will
	see that, with the help of optimization, excellent ground state fidelities
	can be achieved for P1 with total ramping times shorter than in
	{\cite{Fang:2024uyf}}, and even shorter times for P3.
	
	\item (Section \ref{sec:dev}) What is the deviation in our chosen observable (2pt function) due to the prepared state being not exactly the ground state?
	
	\item (Section \ref{sec:decoherence}) What about the decoherence effects?
	The main decoherence effects reduce proportionally to the evolution times.
	Our recipe to reduce them is to reduce the evolution time, while keeping
	ground state fidelity high enough.
\end{enumerate}
Once all of this is discussed, we will summarize the prospects for this measurement as a part of our general conclusions
in Section \ref{sec:concl}.

\subsection{Ground-state 2pt function}\label{sec:2ptDMRG}

In this section we suppose that parameters of the Rydberg chain Hamiltonian
were tuned to lie on the critical line. For concreteness we will consider
points P1 and P3 on this line. We will discuss the 2pt function of microscopic
operators $s_i$ in the ground state, and see how they compare with the CFT 2pt
function of the primary $\sigma$. How to realize, via adiabatic evolution, a
state sufficiently close to the groud state will be discussed in Section
\ref{sec:fidelity} below.

\begin{figure}[h]\centering
	\resizebox{250pt}{!}{\includegraphics{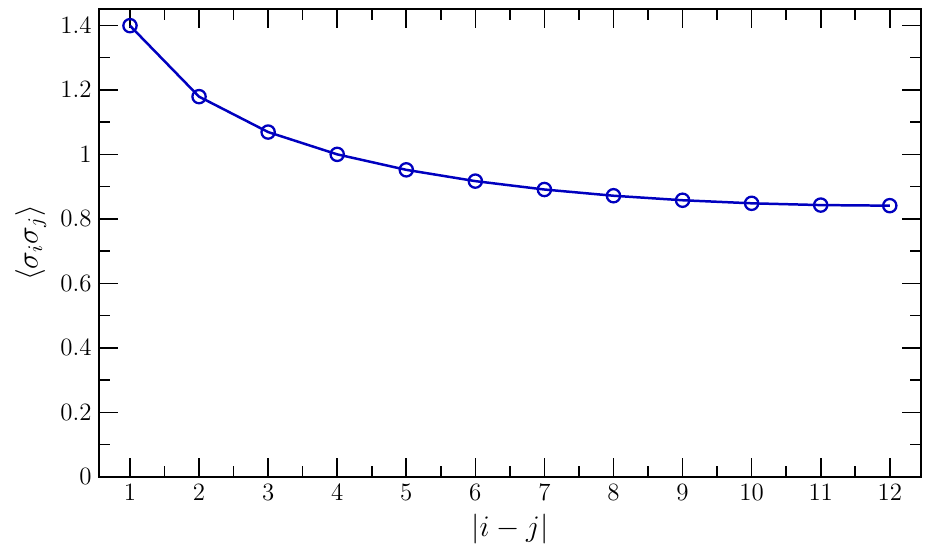}}
	\caption{\label{fig:ideal}CFT prediction for 2pt function at $N = 24$, Eq.
		\eqref{2ptideal}. }
\end{figure}

We map the circle on which Rydberg atoms are positioned to the CFT circle of
length $2 \pi$, so that atom number $i$ is mapped to the angle $\phi_i = 2 \pi
i / N$. If the microscopic operator $s_i$ were exactly equal to the CFT
operator $\sigma_i := \sigma (\phi_i)$,\footnote{We will no longer write time
	for the CFT operators, which is always set to zero.} the ground state
correlation function would be equal to \eqref{eq:2ptcylequal} with
$\Delta_{\mathcal{O}} = \Delta_{\sigma} = 1 / 8$:
\begin{equation}
	{\langle \sigma_i \sigma_j \rangle }  = \frac{1}{\left( 2 \sin \frac{\pi | i
			- j |}{N} \right)^{1 / 4}} . \label{2ptideal}
\end{equation}
This 2pt function is plotted in Fig.~\ref{fig:ideal} for $N = 24$ and for
distance $| i - j | = 1 \ldots 12$.
 This is the dependence that we hope will be seen
experimentally. The range of variation for $N = 24$ is not huge; we have
$\left\langle \sigma_0 {\sigma_{12}}  \right\rangle \approx 0.6 \left\langle
\sigma_0 {\sigma_1}  \right\rangle$.\footnote{It could be suggestive to plot
		\eqref{2ptideal} as a function of \ $\sin \frac{2 \pi i}{N}$ in log-log scale,
		with $\sin \frac{2 \pi i}{N}$ on the horizontal axis, since such a plot would look a straight line. But since the range of variation is not so large, we prefer to use the linear scale as in Fig.~\ref{fig:ideal}.} Still, the shape of the curve is uniquely predicted, and measuring this curve in agreement with the theory will be a direct test of conformality.

Prediction \eqref{2ptideal} is idealized because we can't find a microscopic
operator $s_i$ which exacty equals the CFT operator $\sigma (\phi_i)$. The
best we can hope for is a relation
\begin{equation}
	s_i = \sigma (\phi_i) + \ldots
\end{equation}
where {\ldots} stands for CFT operators of higher scaling dimension than
$\sigma$.

To find an appropriate operator $s_i$, let us start with $Z_i$, the simplest
microscopic operator diagonal in the $n_i$ basis (so that its correlation function is accessible in Rydberg atom experiments). We can expand $Z_i$ in a sum
of $\mathbb{Z}_2$-even and $\mathbb{Z}_2$-odd CFT operators, as follows:
\begin{equation}
	Z_i = (- 1)^i \sum_{\mathbb{Z}_2 \text{-odd } \mathcal{O}} c_{\mathcal{O} }
	\mathcal{O}  (\phi_i) + \sum_{\mathbb{Z}_2 \text{-even } \mathcal{O}}
	c_{\mathcal{O}} \mathcal{O} (\phi_i)
\end{equation}
In the considered lattice model, lattice translation by one site maps to a translation accompanied by the
global $\mathbb{Z}_2$ transformation of the CFT. This explains factor $(- 1)^i$ in
front of the sum over $\mathbb{Z}_2$-odd $\mathcal{O}$.

The first $\mathbb{Z}_2$-odd CFT operator is $\sigma$ that we are after, and
the first $\mathbb{Z}_2$-even CFT operator is the identity, which can be
subtracted by subtracting the vev of $Z_i$. Thus we identify the microscopic
operator \cite{Slagle:2021ene}
\begin{equation}\label{latticeoperator}
	s_i = (- 1)^i (Z_i - \langle Z_i \rangle) .
\end{equation}
We then have that $s_i = c_{\sigma} \sigma (\phi_i)\ +$ corrections from
higher scaling dimension operators:
\begin{equation}
	s_i = c_{\sigma} \sigma (\phi_i) {+ c_{\partial_{\phi}^2 \sigma}} 
	\partial_{\phi}^2 \sigma (\phi_i) + (- 1)^i c_{\varepsilon} \varepsilon
	(\phi_i) + \ldots \label{sicorr}
\end{equation}
where we kept in the RHS the lowest $\mathbb{Z}_2$-odd
and $\mathbb{Z}_2$-even corrections. This is all what can be said in general,
and in particular for point P1. For point P3, because of the onsite $\mathbb{Z}_2$
symmetry, we have simplifications. All $\mathbb{Z}_2$-even operators have
$c_{\mathcal{O}} = 0$. Thus $\langle Z_i \rangle = 0$ so that the expression
for $s_i$ simplifies. Also $c_{\varepsilon} = 0$ so that $\varepsilon
(\phi_i)$ drops out from the correction term in \eqref{sicorr}.

Both for P1 and P3, since the correction terms in \eqref{sicorr} have higher
scaling dimension than the leading term, we expect that the idealized
prediction \eqref{2ptideal} will hold at sufficiently large distances $| i - j
| \gg 1$. At how large distances this holds depends on how
large coefficients $c_{\mathcal{O}}$ of the correction term are with respect
to $c_{\sigma}$. 

To answer this, we pass on to simulations. We numerically compute the ground state
wavefunction for points P1 and P3. The ground state belongs to the
translationally invariant, spatial parity invariant subspace of the full
Hilbert space, which for $N = 24$ has dimension $\mathcal{D}=352, 698$. This is sufficiently small so that for $N = 24$
the ground state and the low-lying energy states can be studied using exact diagonalization (ED), i.e.~the Lanczos method, avoiding the recourse to DMRG. This will be important for the optimization of adiabatic evolution in Section \ref{sec:fidelity}.

\begin{figure}[h]\centering
	\resizebox{0.6\columnwidth}{!}{\includegraphics{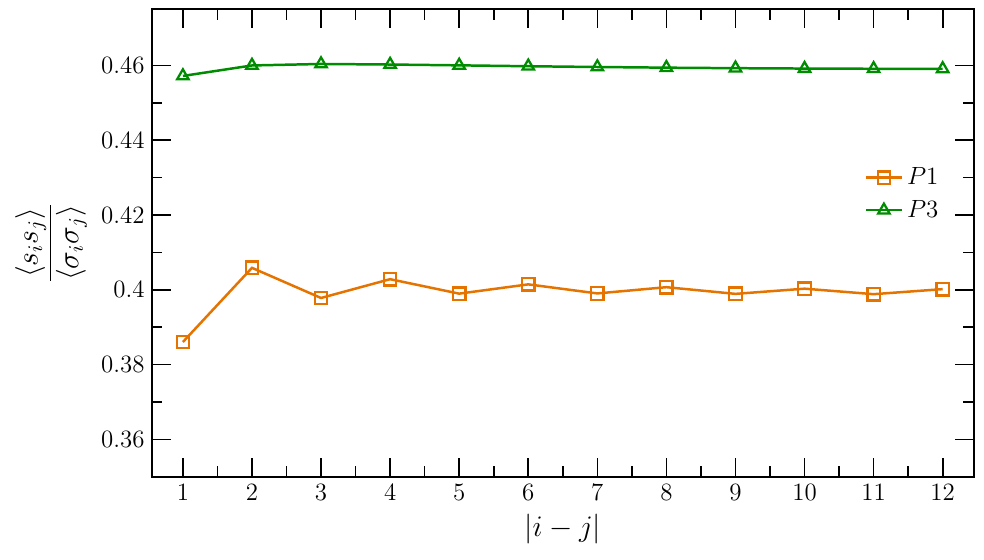}}
	\caption{\label{fig:2pt-sim}The 2pt function for the $N=24$ Rydberg spin chain at criticality (points P1 and P3), divided by the CFT prediction from Fig.~\ref{fig:ideal}. }
\end{figure}

We thus compute the ground state and evaluate the ground state 2pt function $\langle 0| s_i s_j |0 \rangle$, which by the
symmetries of the problem is translationally and spatial parity invariant,
i.e.~depends only on $| i - j |$. The result is shown in Fig.
\ref{fig:2pt-sim}, divided by the CFT prediction \eqref{2ptideal}. We see that
the ratios rapidly asymptote to a constant for both points P1 and P3. The
curves do show deviation from a constant at small $| i - j |$, but these deviations are small (a few $\%$) even for separation 1. These
deviations are due to the correction terms in \eqref{sicorr}. We see that for
point P3, where the correction term $(- 1)^i c_{\varepsilon} \varepsilon
(\phi_i)$ is absent, the approach to a constant is smoother and is monotonic.

Previously, numerical simulations of 2pt functions of the Rydberg spin chain
were discussed in {\cite{Slagle:2021ene}}, focusing on the strong Rydberg
blockade regime where FSS effective Hamiltonian is applicable, and using the
same microscopic operator $s_i$, up to rescaling. There, oscillations in the
2pt function due to admixture of $(- 1)^j \epsilon (\phi)$ were also observed.

\subsection{Estimation of the needed sample size}\label{sec:estimation-of-the-needed-sample-size}

\label{sec:sample}We learned in the previous section that the critical 2pt
function for relatively short chains of Rydberg atoms ($N = 24$) asymptotes to
its CFT form at distances order 1. In this section we will discuss how many
times one has to repeat the experiment to measure this 2pt function.

Consider the Hilbert space basis consisting of states $|n_0, \ldots, n_{N - 1}
\rangle$ where each $n_i = 0, 1$. We can expand the ground state in this
basis:
\begin{equation}
	| 0 \rangle = \sum_{n_i = 0, 1} a_{n_0, \ldots, n_{L - 1}} |n_0, \ldots,
	n_{L - 1} \rangle .
\end{equation}
In the experiment we will not have access to the full ground state
wavefunction itself but to collapsed states (``snapshots'') $\Psi_m$, $m = 1,
\ldots, M$, where $M$ is the number of times we repeated the experiment. Each
$\Psi_m$ is one of the basis states $|n_0, \ldots, n_{L - 1} \rangle$ and they
will appear in the experiment according to the Born rule: the probability to
observe the snapshot $|n_0, \ldots, n_{L - 1} \rangle$ equals $|a_{n_0,
	\ldots, n_{L - 1}} |^2$.

Any observable which is a function of $n_i$'s can be obtained experimentally
averaging over a sufficiently large number of collapsed states. The 2pt
function plotted in Fig.~\ref{fig:2pt-sim} is such an observable. How many
collapsed states is needed for a good signal-to-noise ratio? I.e.~how many
times $M$ shall we have to repeat the experiment? To answer this question, we
need to study the {{\em variance}} of our observable.

In principle, the variance of any observable $\mathcal{O}$ can be computed from the expected
value of $\mathcal{O}^2$. Here we will not compute the variance but will estimate it, using a method
which is more direct and closer to the experiment. First, we compute the
ground state for $N = 24$ using ED, working in the subspace
of translation and parity-invariant states, each of which is a finite linear
combination of  the basis states $|n_0, \ldots, n_{L - 1} \rangle$. We then
sample the ground state, i.e.~generate a sequence of random snapshots $|n_0, \ldots, n_{L - 1} \rangle$ distributed according to the
Born rule.\footnote{This is done in two steps: we first pick a random translation and
parity-invariant state, and then retain one of its constituents $|n_0, \ldots,
n_{L - 1} \rangle.$.} This allows us to
simulate the experimental process directly. Statistical errors will go down as
$\Sigma / \sqrt{M}$, where $M$ is the number of snapshots, with $\Sigma$
estimated from the ``ED data''.\footnote{For sufficiently
	large $N$, ED will become unfeasible and one will have to switch to DMRG {\cite{White1992,Schollwock2011}}. This produces the wavefunction as
	a matrix product state (MPS). Interestingly, MPS representation also allows
	for quick sampling, using the {\texttt{sample()}} function of the
	{\texttt{iTensor.jl}} package.}
	
	\begin{figure}[h]\centering
		\begin{center}
			\resizebox{0.6\columnwidth}{!}{\includegraphics{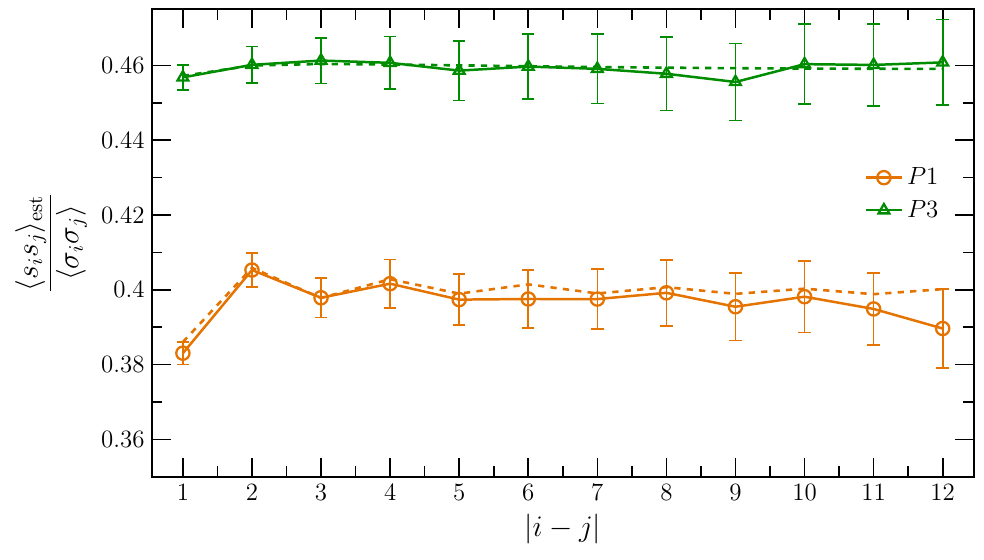}}
		\end{center}
		\caption{\label{ZZ2ptsample}Estimating the critical 2pt function on the ring of $N = 24$ atoms
			from independent snapshots, for points P1 and P3. The dashed lines are the ED results from Fig.~\ref{fig:2pt-sim}. The central values and the error bars are obtained as described in the text, averaging over $M = 2 \times 10^3$ snapshots.}
	\end{figure}

In our calculation we picked $M = 2 \times 10^3$. For each of $M$ collected
snapshots $\Psi_m$ we measure the 2pt function $\langle s_0 s_j
\rangle$ as a function of $j$ (averaging over translations):
\begin{equation}
	\label{sample} X_m (j) = \frac{1}{N}  \sum_{i = 0}^{N - 1} \langle \Psi_m
	|s_i s_{i + j} | \Psi_m \rangle .
\end{equation}
For a fixed $j,$ $\{ X_m (j) \}_{m = 1}^M$ is a set of $M$ independent
identically distributed random variables, whose mean $\overline{X (j)}$ is the
ground-state 2pt function $\langle 0| s_0 s_j |0 \rangle$. The mean and the
standard deviation $\Sigma$ are estimated by the standard formulas (see for example~\cite{Taylor1997ErrorAnalysis}),
\begin{equation}
	\overline{X (j)}_{\rm est} = \frac{1}{M}  \sum_m X_m (j), \qquad \Sigma
	(j)^2_{\rm est} = \frac{1}{M - 1}  \sum_m (X_m (j) - \overline{X (j)})^2
	\hspace{0.17em} .
\end{equation}
By the central limit theorem, for $M \to \infty$, the sample mean $\overline{X
	(j)}_{\rm est}$ has a gaussian distribution with variance $(\Sigma (j) /
\sqrt{M})^2$ around the true mean. We therefore obtain an estimation of the
2pt correlator, including one standard deviation error bars, as
\begin{equation}
	\langle s_0 s_j \rangle_{\rm est} = \overline{X (j)}_{\rm est} \pm
	\frac{1}{\sqrt{M}}  \sqrt{\Sigma (j)^2_{\rm est}} . \label{eq:conf}
\end{equation}
In Fig.~\ref{ZZ2ptsample} we show the result of applying this procedure to the
critical 2pt function for $N = 24$ and $M = 2 \times 10^3$ snapshots. We see
that the statistical error is only a few percent, and the results agree with the ED calculation from Fig.~\ref{fig:2pt-sim} within that error.

\subsection{Approximate ground state preparation}
\label{sec:fidelity}
In the previous sections, we saw that if the
critical ground state can be obtained, then the critical 2pt function can be
measured, with good statistics achievable with $O (10^3)$ repetitions. 

Experimentally, one realizes the critical ground state via adiabatic
evolution.

Let us follow the same protocol as in {\cite{Fang:2024uyf}}, starting the evolution in the disordered phase
with $\Delta' < 0$, $\Omega' = 0$ in the state $| 00 \ldots 0 \rangle$ (i.e.~all atoms in the ground state). This state is the exact ground
state for $\Delta' < 0$, $\Omega' = 0$. One then picks the curve in the $(\Omega',\Delta')$ plane
(called ``ramp profile'') $(\Omega' (t), \Delta' (t))$ which ends at a point on
the Ising critical line. We denote by $T$ the total evolution time. By the adiabatic theorem, if the evolution is
sufficiently slow, the end state will approximate the critical ground state in
finite volume. However, for experimental reasons, evolution cannot be
arbitrarily slow. One wants it to be sufficiently slow so that good ground
state fidelity is achieved, yet as fast as possible, to reduce decoherence
effects to be discussed in Section \ref{sec:decoherence}.

\begin{figure}[h]\centering
	\resizebox{0.55\columnwidth}{!}{\includegraphics{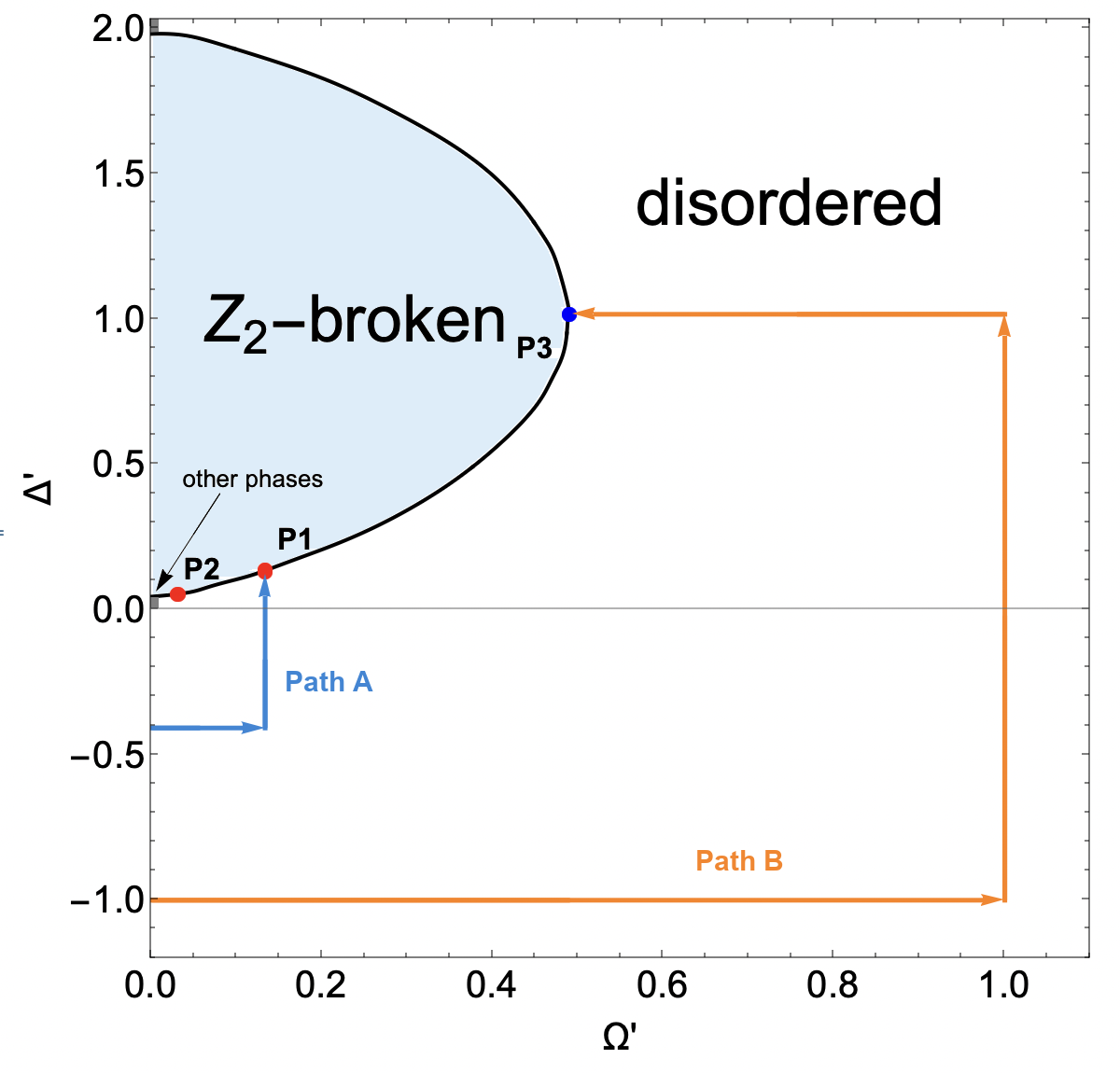}}
	\caption{\label{fig:evolution}Adiabatic evolution paths A and B discussed in this work.}
\end{figure}

For optimization purposes, we will fix a path in the $(\Omega', \Delta')$
plane and we will vary the local rate of change and the total evolution time.
For definiteness, we fix a two-segment Path A for point P1 and a three-segment
Path B for P3, see Fig.~\ref{fig:evolution}.

Path A is precisely the one used in {\cite{Fang:2024uyf}}, where
total evolution time was $T= 7 \mu \text{s}$ split as $2 \mu \text{s} + 5
\mu \text{s}$ between the two segments. Their ramp profile $\Omega' (t),
\Delta' (t)$ is shown in Fig.~\ref{fig:yao-ramp}. In particular, linearly growing
$\Omega' (t)$ was chosen during the first segment, while during the second
segment $\Delta' (t)$ was chosen as a rational function with decreasing first
derivative. This was motivated by the locally adiabatic approximation
{\cite{Roland:2001imn}}, which parametrizes evolution in terms of the gap
between the ground state and the first excited state. The gap closes at the
critical point in the thermodynamic limit, while remaining finite at finite
$N$.

Ref.~\cite{Fang:2024uyf} did not fully optimize their ramp profile, citing high computational cost. The point we want to make here is that 
$N = 24$ is small enough so that efficient optimization is feasible. In fact, we can find better ramp profiles for Path A leading to point P1. Better here means achieving as good or better ground state fidelity than the profile in \cite{Fang:2024uyf} while at the same time having significantly shorter evolution time, which should reduce decoherence effects (see Section \ref{sec:decoherence}). Even better ramp profiles will be found for Path B leading to point P3. Let us describe these results.

First of all, let us describe how to check if a given profile is good enough (``validation''). For $N=24$, and any given ramp profile, we can solve numerically the Schr\"odinger time evolution
\begin{equation}
	i \hbar \dot{\psi} (t) = U H (\Omega' (t), \Delta' (t)) \psi\,, \label{fullQM1}
\end{equation}
starting from the state $| 00 \ldots 0 \rangle$. (We fix $U = \hbar \times
75.69\,\text{MHz}$ as in {\cite{Fang:2024uyf}}, see App.~\ref{sec:Yao-details}.) The point is that this evolution happens in the $\mathcal{D}$-dimensional translation and
spatial-parity invariant subspace of the full Hilbert space to which the ground state belongs, and $\mathcal{D}$ is small enough for $N=24$, so that this computation is feasible;\footnote{All computations in this paper were done in few tens of minutes on a laptop.} see Appendix \ref{sec:validation} for details. 

At any time $t$, the wave function $\psi (t)$ can be expanded in the energy eigenstates of the Hamiltonian at time
$t$:
\begin{equation}
	\psi (t) = \sum_{n = 0}^{\infty} c_n (t)  | n (t)  \rangle .
	\label{eq:psiexp12}
\end{equation}
Note that extracting coefficients $c_n (t)$ requires computing $\psi(t)$ and the eigenstates of $H(t)$. Our method of solving \eqref{fullQM1} does not require diagonalizing $H(t)$ for every $t$, which would be expensive. However, at time $t=T$ we do compute the first few low-lying eigenstates of $H(T)$, which lies on the critical line, and in particular the critical ground state $|0(T)\rangle$.

We refer to $| c_0 (T) |^2$ as the ramp fidelity. The closer it is to 1, the
better the final state approximates the critical ground state. 

In Table \ref{tab:fidel} we give ramp fidelities, as well as $| c_n (T) |^2$ for $n=1,2$, for several ramps to P1 and P3. These numbers give an idea for the size of the deviation from the ground state. 

The first line in Table \ref{tab:fidel} is for the ramp from \cite{Fang:2024uyf}, which we call P1$_{7}$, and for which we find fidelity $99.04\%$. The subsequent lines give several ramps that we found via optimization process described below. The subscript in the ramp name refers to the total ramp time in $\mu{\text{s}}$. All these ramps achieve as good fidelities as P1$_7$, or better, with 2-3 times shorter evolution times for point P1,
and even shorter times for point P3. E.g.~ramp P3$_{1.2}$ achieves $99.9\%$ fidelity with $T=1.2 \mu\text{s}$.

\begin{table}[h]\centering
\begin{tabular}{llccccc}
	\toprule
	& 
	& $T\,(\mu\mathrm{s})$
	& $|c_0(T)|^2$
	& $|c_1(T)|^2$
	& $|c_2(T)|^2$
	& $1-\displaystyle\sum_{n=0}^{2}|c_n(T)|^2$ \\
	\midrule
	ramps to P1 & P1$_7$ \cite{Fang:2024uyf} & $7.0$ & $0.990435$ & $0.004265$ & $0.001305$ & $0.003995$ \\
	& P1$_{2.5}$ & $2.5$ & $0.991810$ & $0.000924$ & $0.005148$ & $0.002117$ \\
	& P1$_3$ & $3.0$ & $0.994031$ & $0.002878$ & $0.001574$ & $0.001517$ \\
	\midrule
	ramps to P3 & P3$_{0.8}$ & $0.8$ & $0.996062$ & $0.000364$ & $0.000975$ & $0.002599$ \\
	& P3$_{1.2}$ & $1.2$ & $0.998985$ & $0.000036$ & $0.000186$ & $0.000793$ \\
	\bottomrule
	\end{tabular}
\caption{\label{tab:fidel}Fidelities for various ramp profiles for points P1 and P3, discussed in the text.}
\end{table}
Let us describe the optimization process (Appendix \ref{sec:reduced}). Let us focus on Path A for definiteness, Path B being completely analogous. We choose a fixed parametrization of the path as $\Omega'(q)$, $\Delta'(q)$, $q\in [0,q_{\max}]$, $q_{\max}=2$. Finding an optimal ramp means finding dependence $q(t)$ so that the fidelity is maximized. In the optimization stage we don't optimize the exact fidelity (which is too costly to compute), but an approximate fidelity. This approximate fidelity, for a given $q(t)$, is computed by integrating an approximate Schr\"odinger equation, obtained by projecting the full Schr\"odinger equation from the $\mathcal{D}$-dimensional space of translation and parity invariant states to a $P$-dimensional subspace spanned, for each $q$, by the lowest $P=20$ energy eigenstates of the Hamiltonian. This projection to the $P$-dimensional subspace depends on the path but not on the function $q(t)$, and it is performed only once, before the optimization is started. Approximate fidelity is therefore feasible to compute thousands of times as is needed for the optimization process. After optimization, the true fidelity is evaluated from the full Schr\"odinger equation; this is the fidelity reported in Table \ref{tab:fidel}.

\begin{figure}[ht]
  \centering
    \includegraphics[width=0.49\columnwidth]{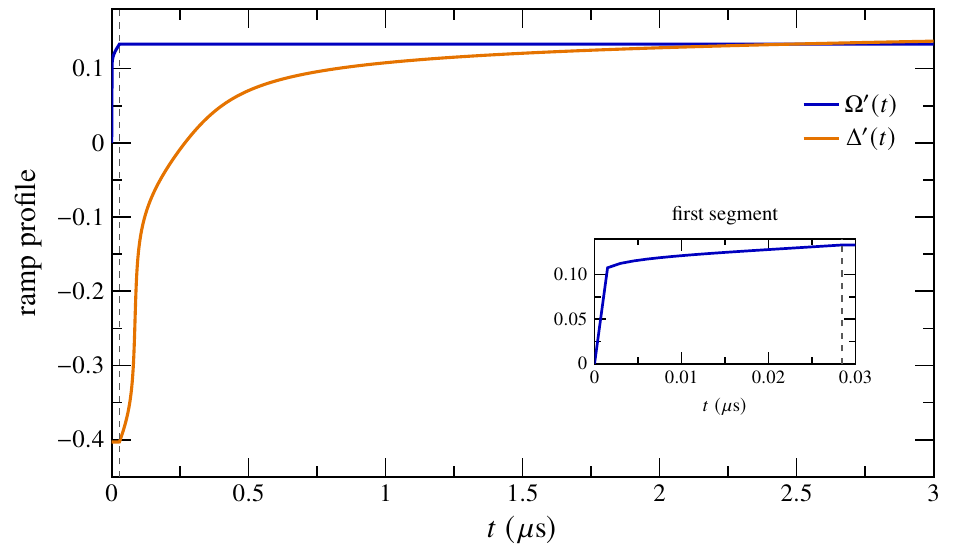}
  \includegraphics[width=0.49\columnwidth]{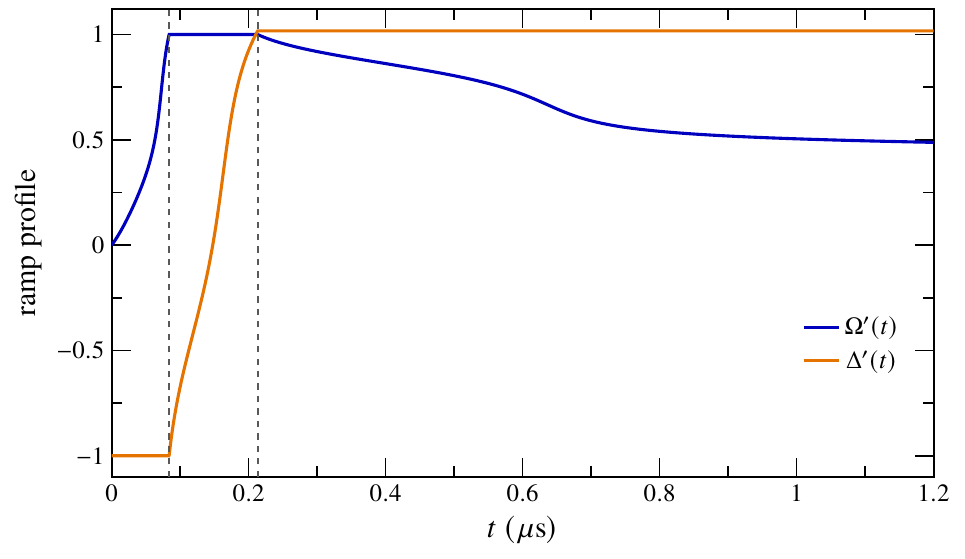}
  \caption{  \label{fig:rampingprofile}Left: the optimized ramp profile P1$_{3}$ with $T=3\mu {\rm s}$.
  Right: The optimized ramp profile P3$_{1.2}$ with $T=1.2\mu {\rm s}$. }
\end{figure}

We see that the four optimized ramp profile given in the table achieve higher fidelity with shorter time than the unoptimized ramp profile used in the experiment \cite{Fang:2024uyf}. For example, if we choose to ramp to P3 along the ramp P3$_{1.2}$, we get $99.9\%$ fidelity with $T=1.2\mu {\rm s}$. The corresponding ramp profile is shown in Fig.~\ref{fig:rampingprofile}, right.
See Appendix \ref{sec:optim} for further details of the optimization algorithm and the optimized ramp profiles for different $T$.

A comment is in order about why the optimized ramps leading to P3 end up having smaller ramp times for the same or better fidelity than the ramps leading to P1. This is due to a combination of two factors. First, we found that the energy gap at P3 is factor $\sim 3$ larger than at P1 for the same $N$ and $U$. This can be explained by noticing that small $\Omega',\Delta'$ at P1 create a new energy scale and suppress the energy gap. Second, we saw that the matrix elements between the ground state and the excited states are also smaller on Path B than on Path A, at least on the part of the path where it approaches the critical line. We believe that this second property is at least partly related to the fact that on the third segment of Path B the Hamiltonian has onsite $\mathbb{Z}_2$ symmetry.

\subsection{Deviation in 2pt function due to imperfect ground state}\label{sec:dev}

\begin{figure}[h]
	\centering
	\includegraphics[width=0.8\columnwidth]{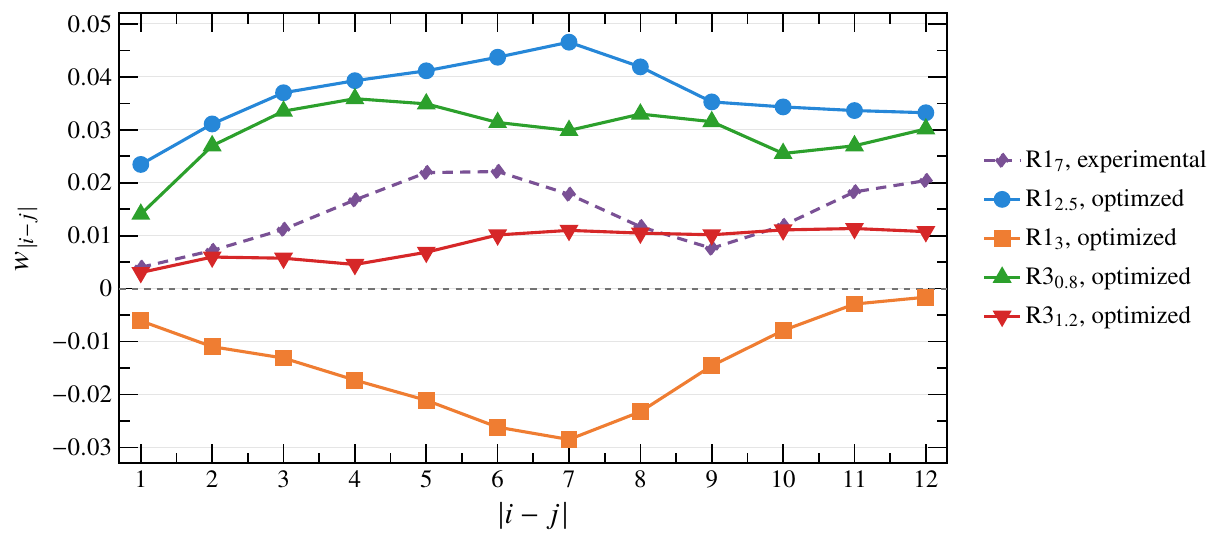}
	\caption{\label{fig:rampingerror}The relative error of the 2pt function, Eq.~\eqref{def:error}, for our optimized ramp profiles and the experimental profile from \cite{Fang:2024uyf}.}
\end{figure}

In the previous section, we explained how to obtain an approximate ground state having fidelity $F$ close to 1. But how close to 1 should it be so that our observable---the 2pt function---stays sufficiently close to the exact ground state prediction? In App.~\ref{sec:dev2ptth} we give a theoretical argument that the deviation scales proportionally to the coefficients $c_1(T)$, $c_2(T)$, $\ldots$, i.e.~it is order $(1-F)^{1/2}$. Thus we expect a few percent deviation due to imperfect fidelity. Here we show a direct check of that prediction. Consider the ratio:
\begin{equation}\label{def:error}
	w_{|i-j|} = \frac{
		\langle \tilde 0| s_i s_{j}|\tilde 0\rangle-
		\langle 0|  s_i s_{j} |0\rangle
	}{
		\langle 0|s_i s_{j}0|\rangle
	}.
\end{equation}
Here $\langle 0|  s_i s_{j} |0\rangle$ is the 2pt function in the exact ground state $|0\rangle$ for the end point of the path, while $\langle \tilde 0|  s_i s_{j} |\tilde 0\rangle$ is the 2pt function in the approximate ground state $|\tilde 0\rangle$ obtained after ramping. Thus $w_{|i-j|}$ is the relative error in the 2pt function after ramping w.r.t. to the exact ground state 2pt function. By the translation and parity invariance, it depends only on $|i-j|$. In Fig.~\ref{fig:rampingerror} we plot this error for all ramp profiles from Table \ref{tab:fidel}. It has the expected size.

\subsection{Decoherence effects}\label{sec:decoherence}

The final issue we need to discuss is ``decoherence''. By this we mean effects which disturb the quantum mechanical evolution preparing the state (e.g.~the atomic motion and Rydberg state decays), as well as effects which perturb the measurement process.
	
\subsubsection{Atomic motion}
By atomic motion we mean deviations of Rydberg atom positions from the time-independent, equally spaced distribution on the circle, shown in Fig.~\ref{fig:chain}. Ref.~\cite{Fang:2024uyf} mentions that atomic motion has non-negligible effect on the observed decoherence correlation length $\xi_{\rm d}$. Here we discuss how smaller ramp time $T$ allows to reduce atomic motion.

Recall that before the quantum-mechanical evolution begins, atoms sit in laser-tweezer trap potential wells. We can assume for simplicity that the position wavefunction of each atom is in the ground states of the potential well, of typical width $\sigma_x$.\footnote{This is an oversimplification because in practice there is an effective temperature and the atom can be in an excited potential well state as well. But in practice the average occupation number $\bar n$ may be quite close to zero. E.g.~\cite{Fang:2024uyf} cites $\bar n\lesssim 0.1$.} In the discussion below we only take into account the atomic displacements along the circle. These are controlled by the ``radial'' trap profile. The ``axial'' trap profile, in the direction orthogonal to the circle plane, is more shallow, but displacements in this direction enter quadratically in the variation of interatomic distances, so this effect is subleading. 

Then, the traps are switched off, and the quantum-mechanical evolution begins (accompanyied by the ramping of $\Omega$ and $\Delta$). Unlike the static, uniformly distributed atomic positions shown in Fig.~\ref{fig:chain}, each atom will have an initial position uncertainty $\sigma_x$ and an initial velocity spread around zero $\sigma_v= \hbar / (2m \sigma_x)$.\footnote{The precise coefficient $1/2$ is assuming the harmonic oscillator ground-state wavefunction.} In plain language, atoms start ``flying away'' from their initial positions. Of course, if the atoms move by a significant amount during the time of the adiabatic evolution, this displacement will affect the strength of the van der Waals interaction between the atoms. So, they better move as little as possible. This requires that either the evolution time or small, or $\sigma_v$ is small. However, reducing $\sigma_v$ requires enlarging $\sigma_x$, which increases uncertainty in the initial atomic positions, thus also influencing the van der Waals interaction strength. The full uncertainty after the evolution time $T=T_{evolve}$ is
\begin{equation}
	\sigma_x^{\rm full} \sim\sigma_x + \frac{\hbar}{2m \sigma_x } T \ge 2 \left(\frac {\hbar T}{2m}\right)^{1/2}\,,
	\label{dxfull}
\end{equation}
where $m$ is the mass of the Rydberg atom. The minimum value of $\sigma_x^{\rm full}$ given in the RHS of \eqref{dxfull} is realized for
\beq
\sigma_x = \left(\frac {\hbar T}{2m}\right)^{1/2}= 15.5\, {\rm nm} \left (\frac{133}{A}\right)^{1/2} \left (\frac T {\mu{\rm  s}}\right )^{1/2},
\label{eq:dxref}
\eeq
where we normalized the atomic weight $A$ to that of ${}^{133}$Cs used in \cite{Fang:2024uyf}. We see that by lowering $T$ from $7\mu {\rm s}$ as in \cite{Fang:2024uyf} to $1.2\mu {\rm s}$ (our optimized ramp R3$_{1.2}$) we can potentially reduce $\sigma_ x^{\rm full}$ by $(7/1.2)^{1/2}\approx 2.4$. Note that the initial radial tweezer trap frequency $\omega_r=2\pi\times 90 {\rm kHz}$ of \cite{Fang:2024uyf} corresponds to position uncertainty along the circle $\sqrt{\hbar/(2 m\omega)}=21\,{\rm nm}$, close to the reference value in \eqref{eq:dxref}. But since \cite{Fang:2024uyf} uses $T=7\mu{\rm s}$ ramp, they adiabatically lowered the radial trap frequency by factor 3 before releasing the atoms. Doing so they balanced the spread due to the second term in \eqref{dxfull}. Our point here is that if a future experiment instead uses the optimized ramp, it can work directly at the initial trap frequency, thus reducing the atomic motion.

\subsubsection{Rydberg state decays}

The Rydberg state decays to the ground state with the lifetime $1/\gamma_{\rm decay}=71.44\,\mu{\rm s}$ \cite{Fang:2024uyf}. Over the course of a ramp time $T$, a Rydberg state atom has decay probability $P_{\rm decay}=1-e^{-\gamma_{\rm decay} T}$ which is $9.3\%$ for $T=7\,\mu{\rm s}$. This effect is therefore a significant contribution to decoherence \cite{Fang:2024uyf}, even though an atom is not fully in the Rydberg state for the whole ramp.

Intermediate-state scattering decays, characterized by a similar timescale, represent another effect of comparable significance. These two effects are modeled in \cite{Fang:2024uyf} using Lindbladian dynamics. While we do not repeat these simulations here, we note that the influence of these two effects scales linearly with $T$, and it will be reduced accordingly when using the optimized ramps.

\subsubsection{Rydberg detection perturbations}

As mentioned, at the end of the adiabatic evolution, projective measurement is performed. This is done by a rapid turn on of the tweezer traps, expelling atoms in the Rydberg state, recapturing atoms in the ground state, and subsequent imaging, to determine the population of each site. Ref.~\cite{Fang:2024uyf} cites $2\%$ probability that a ground state is misidentified as a Rydberg state (``false positive'') in the process, and $<1.5\%$ probability for the opposite misidentification (``false negative''). These probabilities being smaller than the above-mentioned Rydberg state decay probability $P_{\rm decay}=9.3\%$ to which Ref.~\cite{Fang:2024uyf} was subjected, presumably they represented a subdominant source of error in the decoherence budget of \cite{Fang:2024uyf}.

Ref.~\cite{Fang:2024uyf} relied on post-selection to eliminate snapshots with blockade violation (two neighboring atoms in the Rydberg state), as a way to mitigate the above ``false positives''. That was a possible strategy as they worked at point P1 which is in the near-Rydberg blockade regime. (It should be noticed that this strategy creates a bias, since about $8\%$ of true snapshots at P1 do contain blockade violations \cite{Fang:2024uyf} and are also eliminated.)  It remains to be seen if measurements can be reliably performed at point P3, which is far away from Rydberg blockade regime, and the above post-selection strategy is not feasible.

We are hopeful that this should be possible, also because the ``false positive'' probability is at least partly related to the atomic motion: if an atom has moved too far from the trap center, it is hard to recapture. This probability is expected to decrease for shorter ramp times $T$. E.g.~\cite[Extended Data Fig. 2]{Bernien:2017ubn} shows that a shorter trap-off time leads to a higher detection rate.

\section{Summary and conclusions}\label{sec:concl}

The main purpose of this paper was to attract attention to direct experimental tests of conformality in critical phenomena. Concretely, we chose to discuss here quantum critical phenomena realized in a currently very popular experimental platform---Rydberg atoms. Even more concretely, we focused on one observable---equal-time 2pt function in a periodic chain of Rydberg atoms at a critical point in the Ising universality class. Previously, this 2pt function was experimentally studied in \cite{Fang:2024uyf}, who saw short-distance behavior consistent with CFT, although longer-distance behavior was modified with respect to CFT by decoherence effects, precluding a convincing direct test of conformality.

In Section \ref{sec:prop}, we went over the main steps of the experimental measurement, having in mind to understand what it would take to achieve say a $10\%$ agreement with theory in a wide range of distances. We did all our computations for two points on the critical line. The first point (P1) is the one used in \cite{Fang:2024uyf} (up to a small but non-negligible correction in the detuning parameter $\Delta$, which sets the point right on the critical line as opposed to just near it). This point sits near the Rydberg-blockade regime, where there was a lot of theoretical work using the effective FSS model \cite{Slagle:2021ene,FSS}. The second point (P3) sits at the tip of the $\mathbb{Z}_2$-broken region of the phase diagram and is as far away from the Rydberg blockade regime as possible. It has onsite $\mathbb{Z}_2$ symmetry. We felt that this region of the critical line could offer advantages with respect to the Rydberg-blockade region, and this turned out indeed the case, see below.

Our main findings are as follows.
\begin{itemize}
	\item
	(Section \ref{sec:2ptDMRG}) Working on the circular chain of $N=24$ atoms, we saw that the exact ground state 2pt function agrees well with the CFT prediction, starting from distances $|i-j|=1$. The agreement is good for both points P1 and P3, with maximal error about $4\%$ for P1 and less that $1\%$ for P3, see Fig.~\ref{fig:2pt-sim}.
	\item
	(Section \ref{sec:estimation-of-the-needed-sample-size}) The variance of the ground state 2pt function is sufficiently small so that good statistics can be obtained with order $10^3$ repetitions of the experiment. Concretely, for $2\times 10^3$ repetitions we find statistical errors $<2.5\%$ for both P1 and P3, see Fig.~\ref{ZZ2ptsample}.
	\item
	(Section \ref{sec:fidelity}) Approximate ground states with fidelity $>99\%$ can be prepared via quasi-adiabatic evolution with ramp times as short as $2.5\mu{\rm s}$ for P1 and $0.8\mu{\rm s}$ for P3. This requires ramp optimization, feasible with modest computation resources (App.~\ref{sec:optim}). This observation gives us hope that the experiment of \cite{Fang:2024uyf} can be improved. Indeed \cite{Fang:2024uyf} used a suboptimal ramp with a significantly longer evolution time of $7\mu{\rm s}$.
	
	It is in the much shorter ramp time that the advantages of point P3 vs P1 become most clear. Such shorter ramp times are possible for P3 because of the larger overall scale of the energy gap, accompanied by the smallness of the pertinent matrix elements.
	
	\item (Section \ref{sec:dev}) We also check explicitly that the error in the 2pt function due to using an approximate ground state vs the exact one is small for our optimized ramps, $\sim 4\%$ or even less, depending on the ramp, see Fig.~\ref{fig:rampingerror}.
	
	\item (Section \ref{sec:decoherence}) While the detailed modeling of decoherence effects is beyond the scope of this work (see \cite{Dalyac:2026jpu} for a recent comprehensive discussion), we note that reducing the ramp time has clear mitigating effect on many of them. For example, the Rydberg to ground state decays, one of the main contribution to decoherence in \cite{Fang:2024uyf}, will be suppressed linearly in the ramp time. 
\end{itemize}

Based on these investigations, we are fairly certain that repeating the experiment of \cite{Fang:2024uyf} using our optimized ramps will give a strictly better result for the 2pt function. The decoherence length $\xi_{\rm d}$ should become larger, and agreement between experiment and CFT will be achieved in a wider range of distances. We hope that such an experiment will be done, and that it will become the first direct test of conformality from correlator measurements.

\section*{Acknowledgements}

We thank colleagues at the Institut d'Optique, Ecole Polytechnique (Antoine
Browaeys, Thierry Lahaye, David Clement, Thomas Chalopin), PASQUAL (Guillaume
Villaret, Adrien Signoles, Alexander Dauphin), and Institute for Advanced
Study, Tsinghua University (Wenlan Chen, Yingfei Gu, Chengshu Li) for
discussions of cold atom experiments. In particular we thank Guillaume Villaret for sharing and discussing an early draft of \cite{Dalyac:2026jpu}. SR also thanks Hans Peter B{\"u}chler
and Enrico Rinaldi for discussions. We thank Antonio Antunes, Victor Gorbenko, Julio Parra-Martinez, Shu-Heng Shao and Balt van Rees for a discussion of operator matching. SR is partially supported by the Simons
Collaboration on the Probabilistic Paths to Quantum Field Theory (award
SFI-MPS-PP-00012621-16).
JR is supported by the European Union (ERC, QFTinAdS, project number 101087025, PI Balt van Rees). Views and opinions expressed are, however, those of the authors only and do not necessarily reflect those of the European Union or the European Research Council Executive Agency.

\section*{AI use and data availability statement}

Numerical simulations (exact diagonalization, ramp profile optimization) were performed using AI-generated codes, with human supervision and cross-checks. These codes are available from the authors upon request. At the current level of technology, they can be also be easily generated from scratch, given the text of our paper.

\appendix

\section{Some details about the experiment \cite{Fang:2024uyf}}
\label{sec:Yao-details}

Ref.~{\cite{Fang:2024uyf}} measured 1D and 2D arrays of Rydberg atoms tuned to
the quantum critical point in the Ising universality class. We will report here
parameters of their experiments, focusing on the 1D case, relevant for our
paper.

\begin{figure}[h]\centering
	\includegraphics{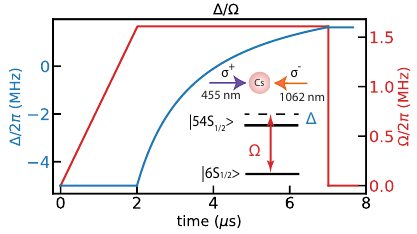}
	\caption{\label{fig:yao-ramp}Evolution $\Delta (t)$ and $\Omega (t)$ (``ramp profile'') for
		preparing the critical state in {\cite{Fang:2024uyf}}, Fig. 1C. }
\end{figure}

The Hamiltonian of {\cite{Fang:2024uyf}} is precisely \eqref{eq:model}
with $N = 24$ atoms on a circle (they also studied $N = 40$). To reach the
critical state they evolve $\Delta, \Omega$ as in Fig.~\ref{fig:yao-ramp}.
They report the following endpoint parameters:
\begin{equation}
	\Omega  / 2 \pi = 1.6\,\text{MHz}, \quad \Delta / \Omega 
	= 0.97 (5), \quad R_b / a \approx 1.4,
\end{equation}
where $R_b = (C_6 / \hbar \Omega)^{1 / 6}$ is the blockade radius (they don't
report the error on $R_b / a$). Combining these we compute:
\begin{equation}
	U = C_6 / a^6 \approx \hbar \times 2 \pi\, 1.6\, 1.4^4\,
{\rm MHz}= \hbar \times
	75.69\,{\rm MHz},
\end{equation}
\begin{equation}
	\Omega' = \hbar \Omega / U \approx 1 / 1.4^6 \approx 0.133 .
\end{equation}
The point P1 in Table \ref{tab:params} has $\Delta' / \Omega' = 1.03$. This is at the
upper end of the range $\Delta / \Omega  = 0.97 (5)$ from
{\cite{Fang:2024uyf}}. Taking instead their central value results in a non-negligible deviation from criticality, as we now discuss.

Ref.~{\cite{Fang:2024uyf}} used the experimental definition of the critical
point, as the location of the maximum of susceptibility $\partial \langle n
\rangle / \partial \Delta$ for sufficiently slow sweeps. They note that this
also coincides with the excitation gap minimum. For measuring the 2pt function
a better theoretical definition of the critical point is where the properties
of the system are described by the CFT Hamiltonian. This may differ by $1 / N$
effects from the definitions used in Ref.~{\cite{Fang:2024uyf}}. Our
simulations show that this deviation is not negligible. Indeed point P1 with $\Delta' / \Omega' = 1.03$ gives the 2pt
function consistent with CFT, while $\Delta' / \Omega' = 0.97$ would give a
2pt function decaying faster than in CFT (thus in the disordered phase), with
$\sim 20\%$ suppression at the largest separation. We therefore use point P1. Also, we advocate that future experiments use the measurement of the 2pt function as a proxy for the criticality, rather than the susceptibility peak.

The 2pt function measured in {\cite{Fang:2024uyf}}, Fig. 2F, was fitted by
them by the curve
\begin{equation}
	\langle s_0 s_j \rangle \propto \delta^{- 1 / 4}_j \exp (- \delta_j / \xi_{\rm d} ),
	\quad \delta_j = a \frac{N}{\pi} \sin \left( \frac{\pi j}{N} \right),
\end{equation}
and $\xi_{\rm d} / a = 13.2^{+ 5.7}_{- 3.6}$ is the correlation length, attributed to
decoherence. The central $\xi_{\rm d}$ value corresponds to $\sim50\%$ suppression of
the correlator at the largest separation w.r.t.~the CFT.

\section{Ramp profile optimization}\label{sec:optim}

We parametrize Path A (see Fig.~\ref{fig:evolution}) as:
\begin{align}
0\leq q\leq1:&\quad
\Omega'(q)=0.133q,\qquad \Delta'(q)=-0.403, \nonumber\\
1\leq q\leq2:&\quad
\Omega'(q)=0.133,\qquad \Delta'(q)=-0.403+0.54(q-1). \label{eq:PA}
\end{align}
The final point is therefore P1: $(\Omega',\Delta')=(0.133,0.137)$, which is critical.
We take Path B to be 
\begin{align}\label{PathB}
0\leq q\leq1:&\quad (\Omega',\Delta')=(q,-1), \nonumber\\
1\leq q\leq2:&\quad (\Omega',\Delta')=\bigl(1,-1+(\zeta(6)+1)(q-1)\bigr), \nonumber\\
2\leq q\leq3:&\quad (\Omega',\Delta')=\bigl(1-0.512(q-2),\zeta(6)\bigr). 
\end{align}
The final point is P3: $(\Omega',\Delta')=(0.488,\zeta(6))$, also a critical point.

With paths so fixed, optimizing the ramp profile means finding a $q(t)$ so that the fidelity at $t=T$ (total ramp duration) is maximized. The optimization algorithm includes 1) optimization in a reduced 20-dimensional
subspace and 2) validation in the full $\mathcal{D}$-dimensional subspace. We discuss validation first.

\subsection{Validation}\label{sec:validation}
We use a matrix-free approach to calculate the time evolution of the ground state.  At $N=24$, the dimension of the translationally invariant, reflection-positive subspace of the Hilbert space is $\mathcal{D}=352{,}698$. (The $\mathcal{D}\times \mathcal{D}$ Hamiltonian is never explicitly constructed and stored.) The Schr\"odinger equation is given in \eqref{fullQM1}. In this appendix, we take $\hat{t} =t\cdot U/\hbar$ to be the dimensionless time. We will drop the hat to simplify the notation. We thus work with the equation
\begin{equation}
i\frac{\partial}{\partial t}\lvert\psi(t)\rangle
=
H(t)\lvert\psi(t)\rangle .\label{eq:Schr11}
\end{equation}
This is formally solved by
\begin{equation}
\lvert\psi(t)\rangle
=
\mathcal{U}(t,0)\lvert\psi(0)\rangle ,
\end{equation}
with the time evolution operator,
\begin{equation}
\mathcal{U}(t,0)
=
\mathcal T
\exp\left[
-i\int_0^t H(t')\,dt'
\right].
\end{equation}
Time-ordering \(\mathcal T\) is necessary because, in general,
$
[H(t_1),H(t_2)]\neq 0.
$

Numerically the time evolution is realized by discretizing the time interval $[0,T]$ into small steps $h=\delta t$, and approximating the evolution operator by a product of matrix exponentials corresponding to small steps. It pays off to organize this discretization efficiently, so that the error scales with a high power $h$, as this reduces the number of time steps needed to achieve a converged result. Here we used the fourth-order commutator-free Magnus (CF4) scheme, whose error is $O(h^4)$~\cite{BlanesMoan2006,AlvermannFehske2011}.

The idea of CF4 is to sample the time-dependent Hamiltonian at two Gauss-Legendre nodes inside the interval:
\begin{equation}
t_{n,1}=t_n+c_1 h,\qquad t_{n,2}=t_n+c_2 h\,.
\end{equation}
Let
\begin{equation}
H_1=H\!\left(\Omega(t_{n,1}),\Delta(t_{n,1})\right),\qquad
H_2=H\!\left(\Omega(t_{n,2}),\Delta(t_{n,2})\right).
\end{equation}
The CF4 scheme defines
\begin{equation}\label{eq:cf4}
 \vert \psi_{n+1}\rangle=
 e^{-i h(a_1 H_1+a_2 H_2)}
 e^{-i h (a_2H_1+a_1H_2)} 
\vert \psi_n\rangle,
\end{equation}
with 
\begin{equation}
c_{1,2}=\frac12\mp\frac{\sqrt3}{6}, \quad
a_{1,2}=\frac14\mp\frac{\sqrt3}{6}.
\end{equation}
It can be shown that Eq.~\eqref{eq:cf4} has error $O(\delta t^5)$ with respect to the exact Schr\"odinger evolution. This single step error leads to the $O(h^4)$ for the full evolution which involves $T/h$ steps.

To use Eq.~\eqref{eq:cf4} we need to be able to efficiently act with a matrix exponential on a state. We discuss this problem in general. Suppose we want to act with $e^{-i h G}$ on a state $|v\rangle$. Forming the full matrix exponential is expensive.
Instead, we form the order-$m$ Krylov space
\begin{equation}
\mathcal K_m(G,v)=\operatorname{span}\left\{
\vert v\rangle,G\vert v\rangle,\ldots,G^{m-1}\vert v\rangle
\right\},
\label{eq:krylovspace}
\end{equation}
tailored to the power series of the exponential $e^{-i h G}\vert v\rangle$, and calculated in a matrix-free manner. 

For Hermitian $G$ as needed for us, the Lanczos iteration constructs an orthonormal basis
\begin{equation}
V_m=(\vert v_1\rangle,\ldots,\vert v_m\rangle),
\qquad
\vert v_1\rangle={\vert v\rangle}/{\lVert v\rVert},
\end{equation}
using the three-term recurrence \cite{Lanczos1950}
\begin{equation}
G\vert v_j\rangle
=\beta_{j-1}\vert v_{j-1}\rangle
+\alpha_j\vert v_j\rangle
+\beta_j\vert v_{j+1}\rangle.
\label{eq:lanczosrecurrence}
\end{equation}
Consequently, the projected representation of $G$,
\begin{equation}
T_m=V_m^\dagger G V_m
=
\begin{pmatrix}
\alpha_1&\beta_1&&0\\
\beta_1&\alpha_2&\ddots&\\
&\ddots&\ddots&\beta_{m-1}\\
0&&\beta_{m-1}&\alpha_m
\end{pmatrix},
\label{eq:tridiagonal}
\end{equation}
is an $m\times m$ real symmetric tridiagonal matrix. 
The exponential action is then approximated as \cite{HochbruckLubich1997}
\begin{equation}
e^{-i h G}\vert v\rangle
\approx 
\lVert v\rVert V_m e^{-i h T_m} \mathbf e_1,
\qquad
\mathbf e_1=(1,0,\ldots,0)^T.
\label{eq:krylovexp}
\end{equation}
Thus, only the small $m\times m$ matrix $T_m$ needs to be exponentiated.  Constructing this matrix requires repeated products $G\vert w\rangle$.\footnote{Since in our calculation $G$ is a linear combination of $H_1$ and $H_2$, we evaluate such products as weighted sum of the previously computed matrix-free actions $H_1\vert w\rangle$ and $H_2\vert w\rangle$.} We stress that neither Hamiltonian, nor generators $G$, nor exponentials are stored as full $\mathcal{D}\times \mathcal{D}$ matrices.  Krylov approximations to matrix-exponential actions and their convergence are analyzed in Ref.~\cite{HochbruckLubich1997}.

Once the full time evolution is completed, we further reduce the remaining $O(h^4)$ error using the Richardson extrapolation \cite{Richardson1911}.  Let $F_h$ denote the final ground-state fidelity obtained for a fixed ramp duration $T$ using CF4 step $h$.  We successively decrease the time step until the error starts scaling as $h^4$. The result is then approximately
\begin{equation}
F_{h}\approx F_{0}+C h^4,
\end{equation}
Here $F_{0}$ is the fidelity obtained in the zero-numerical-step-size limit.  Then, computing $F_h$ and $F_{h/2}$ allows us to estimate $ C$ and to find $F_{0}$.

As a first application, we use this method to analyze the ramp profile P1$_7$ from \cite{Fang:2024uyf}, plotted in Fig.~\ref{fig:yao-ramp}.
In terms of Path A parametrization \eqref{eq:PA}, this ramp profile corresponds to 
\begin{equation}
q(t)=
\begin{cases}
\dfrac{t}{151.38},
&0\leq t\leq151.38,\\[10pt]
1+
\dfrac{
0.453893\,(t-151.38)
}{
0.453893\,(t-151.38)
+
0.030395\,(529.83-t)
},
&151.38\leq t\leq529.83.
\end{cases}
\end{equation}
After the time evolution, we get the fidelity shown in the first line of Table~\ref{tab:fidel}. 

\subsection{Optimization within reduced subspace}\label{sec:reduced}

In the previous subsection we described how, given a ramp profile, Schr\"odinger evolution within the $\mathcal{D}$-dimensional subspace can be performed, computing the fidelity at time $T$. In this subsection we discuss how to optimize the ramp profile. While optimizing, final fidelity will have to be computed thousands of times. Computing the full evolution thousands of times is not feasible. We will therefore truncate the Hamiltonian to a $20\times 20$ matrix. This will introduce an error in fidelity. However this approximation still captures the needed profile shape. Once the optimization is over, we compute accurate fidelities as described in the validation section, and it turns out that when approximate fidelity is close to 1, accurate fidelity is also close to 1. Let us describe this in detail.

Optimizing the ramp profile means finding $q (t)$ such that the final state
achieves high fidelity. We first discuss how we quickly evaluate fidelity for any given $q(t)$.
We start with the Schr{\"o}dinger equation
\eqref{eq:Schr11} which we write as
\begin{equation}
	i \dot{\psi} = H (q (t)) \psi (t)\,,
\end{equation}
to show explicitly how $q (t)$ enters in the problem. We project this equation
on the lowest $P = 20$ eigenstates. Let's discuss this first as if $q$ were
continuous, but eventually $q$ will be discretized. Consider for each $q\in [0,q_{\max}]$ ($q_{\max}=2$ for Path A, $3$ for Path B), the first $P$ eigenstates:
\begin{equation}
	Y  (q) = \left(\begin{array}{c}
		|Y_a (q ) \rangle
	\end{array}\right)_{a = 0 \ldots P - 1}  .
\end{equation}
of energies $(E_a)_{a = 0 \ldots P - 1}$. Thus for every $q$
\begin{equation}
	Y   \in \mathbb{R}^{\mathcal{D} \times P}, \qquad (Y ^{})^T Y  = I_{P\times P} .
\end{equation}
We assume that $Y_a$ are chosen orthonormal and real (since the Hamiltonian is
real), and that $Y$ varies smoothly with $q$. We write the time-dependent
state projected to the $P$-dimensional subspace as:
\begin{equation}
	| \psi (t) \rangle = \vec{c} (t) \cdot \vec{Y} (q (t)) .
\end{equation}
Substituting this into the Schr{\"o}dinger equation we find
\begin{equation}
	i \frac{dc_a}{dt} = E_a (q (t)) c_a (t) - \dot{q} (t)  \hspace{0.17em}
	\sum_b \mathcal{A} (Y)_{ab} (q (t)) c_b (t),
\end{equation}
with the Berry connection
\begin{equation}
	\mathcal{A} (Y)_{ab} (q) = i \langle Y_a (q) | \partial_q Y_b (q) \rangle\,.
\end{equation}

Following {\cite{Kato1950}}, we find it convenient to pass, by an orthogonal
rotation, to another real orthonormal basis $U (q)$ of the same
$P$-dimensional subspace such that the Berry connection $\mathcal{A} (U)$ is
zero. In that basis the Schr{\"o}dinger equation becomes
\begin{equation}
	i \frac{dc_a}{dt} = H_{\text{trun}} (q (t))
	c_a (t), \label{SchrHU}
\end{equation}
where
\begin{equation}
	H_{\mathrm{trun}} (q)_{ab} = \langle U_a (q) |H (q) |U_b (q) \rangle .
	\label{HU}
\end{equation}
Let us now describe how the basis $U (q)$ is found, in a discretized setting
which is of interest to us. We distribute $O(\text{few hundred})$ sample points
$q_j$ along the ramp path $(\Omega' (q), \Delta' (q))$.  At every sampled
point, an iterative eigensolver computes the lowest 20 eigenstates. We thus
compute $Y_j : = Y (q_j)$. We ensure that the variation is smooth, or at least
continuous, by imposing the condition $\langle Y_a (q_j) |Y_a (q_{j + 1})
\rangle > 0$.

We then construct $U_j$ inductively, starting from $U_0 = Y_0$. Given $U_j$,
we seek the basis $U_{j + 1}$ at the next time step in the form
\begin{equation}
	U_{j + 1} = Y_{j + 1} R_j,
\end{equation}
where $R_j$ is an orthogonal rotation matrix. Consider the matrix $M = U_j^T
Y_{j + 1}$, i.e.
\begin{equation}
	M_{a, b} = \langle U_a (q_j) |Y_b (q_{j + 1}) \rangle .
\end{equation}
Perform a singular value decomposition
\begin{equation}
	M_j = U_j^T Y_{j + 1} = W_j \Sigma_j V_j^T,
\end{equation}
where $W, V$ are orthogonal and $\Sigma$ diagonal. We claim that the
orthogonal rotation
\begin{equation}
	R_j = V_j W_j^T . \label{eq:Rj}
\end{equation}
does the job. The discretized vanishing Berry phase condition that we need to
check is
\begin{equation}
	U_j^T \Delta_j = 0, \quad \Delta_j := U_{j + 1} - U_j .
\end{equation}
As we will see, this will not be exactly true, but approximately true, modulo
$\Delta_j^T \Delta_j $ terms, which are small for small $q$ discretization
steps.

First of all we note that for $R_j$ as in \eqref{eq:Rj} we have
\begin{equation}
	U_j^T U_{j + 1} = U_j^T Y_{j + 1} R_j = W_j \Sigma_j V_j^T V_j W_j^T = W_j
	\Sigma_j W_j^T
\end{equation}
is symmetric. Since
\begin{equation}
	U_j^T U_{j + 1} = I + U_j^T \Delta_j,
\end{equation}
this implies that $U_j^T \Delta_j$ is symmetric. On the other hand,
substituting $U_{j + 1} = U_j + \Delta_j$ into $U_{j + 1}^T U_{j + 1} = I$, we
have
\begin{equation}
	\Delta_j^T U_j + U_j^T \Delta_j + \Delta_j^T \Delta_j = 0,
\end{equation}
which means that $U_j^T \Delta_j$ is antisymmetric, modulo $\Delta_j^T
\Delta_j $. Since $U_j^T \Delta_j$ is both symmetric and antisymmetric, it is
zero (modulo $\Delta_j^T \Delta_j $). The described sequence of steps is the same as
for solving the so called orthogonal Procrustes problem
{\cite{Schonemann1966,LiuVanderbilt2015}}.

In practice, we find that the matrix elements of the Hamiltonian \eqref{HU}
constructed with respect to a sufficiently finely spaced grid of $q_j$ points
vary smoothly with $q$. So we define $H_{\rm trun}
(q)$ in between the grid points by interpolation. Note that the described computation for $H_{\rm trun}
(q)$
is performed once for Path A and once for Path B, before starting the
optimization process, so it's not overly costly. With so defined
$H_{\rm trun} (q)$, and for any given $q (t)$, we
can solve \eqref{SchrHU} using the same CF4 scheme as described in App.~\ref{sec:validation}, and compute the fidelity. Because everything is now happening in the $P$-dimensional space, it is feasible to perform this computation thousands of times, as is needed for
the optimization.

We now proceed with describing the optimization details. We parametrize not
$q (t)$ directly, but the function $\rho (q) \propto d t / d q$. Namely, we
use the following ansatz
\begin{equation}
	\rho (q ; \vec{a}, \vec{b}) = \rho_0 (q) \exp \hspace{-0.17em} \left[
	\sum_{k = 1}^{K_s} a_k \sin \hspace{-0.17em} \left( \frac{k \pi q}{q_{\max}}
	\right) + \sum_{k = 1}^{K_c} b_k \cos \hspace{-0.17em} \left( \frac{k \pi
		q}{q_{\max}} \right) \right], \label{eq:density}
\end{equation}
with a fixed function $\rho_0 (q)$ (to be described shortly), and real
parameters $\vec{a}$ and $\vec{b}$, which are adjusted to optimize the ramp
profile. We take $K_s = K_c = 6$, i.e.~the total of 12 parameters.

Given $\rho$ and the total evolution time $T$, the map from time $t$ to the
coordinate $q$ is obtained by solving the equation
\begin{equation}
	u (q) = t / T, \quad u (q) : = \frac{\int_0^q \rho (s) \hspace{0.17em} d
		s}{\int_0^{q_{\max}} \rho (s) \hspace{0.17em} d s}, \label{eq:clock1}
\end{equation}
i.e.
\begin{equation}
	q (t) = u^{- 1}  (t / T) . \qquad \label{eq:clock}
\end{equation}
Note that $u \in [0, 1]$ has the meaning of the normalized time.

For $\rho_0$, we choose
\begin{align}
	\text{Path A: } & \rho_0 (q) = \left\{\begin{array}{ll}
		2, & 0 \leq q \leq 1,\\
		5, & 1 < q \leq 2,
	\end{array}\right. \nonumber\\
	\text{Path B: } &\rho_0 (q) \equiv 1, \label{rho0B}
\end{align}
I.e.~for Path A the zero-coefficient profile preserves the $2 : 5$
segment-time ratio of the experimental ramp.

The optimization problem is to maximize the fidelity, i.e.~the squared overlap
of $\psi (T)$ with the ground state $|0 \rangle$ at $q = q_{\max}$:
\begin{equation}
	\tilde F  = | \langle 0 (q_{\max}) | \psi  (T) \rangle |^2 \rightarrow \max .
	\label{eq:f20}
\end{equation}
We put a tilde on $F$ to emphasize that this is an approximate fidelity, computed using the evolution in the truncated Hilbert space.

\begin{table}[h]
	\centering
	\resizebox{\textwidth}{!}{%
		\begin{tabular}{c*{12}{r}}
			\toprule
			$T~(\mu{\rm s})$
			& $a_1$ & $a_2$ & $a_3$ & $a_4$ & $a_5$ & $a_6$
			& $b_1$ & $b_2$ & $b_3$ & $b_4$ & $b_5$ & $b_6$ \\
			\midrule
			2.00
			&  0.565161 & -1.657209 &  3.299559 &  2.171572 & -1.762571 &  2.607463
			& -3.832125 & -2.438523 & -5.717325 &  3.100494 & -1.826421 & -3.311279 \\
			2.25
			&  0.553266 & -0.634082 &  3.273120 &  4.607602 & -2.649089 &  0.498309
			& -5.271103 & -2.026259 & -5.404912 &  3.881643 &  0.612730 & -3.308933 \\
			2.50
			& -0.096198 & -0.819855 &  2.688715 &  4.455646 & -2.987238 & -0.163164
			& -5.604367 & -2.128336 & -4.931697 &  4.193614 &  1.601918 & -2.771219 \\
			2.75
			& -2.188016 & -0.907333 &  2.860120 &  4.814655 & -3.037231 & -0.667492
			& -4.404842 & -2.495907 & -4.589297 &  4.883557 &  3.250500 & -1.941580 \\
			3.00
			& -0.699069 & -0.863736 &  2.237207 &  6.867565 & -1.721453 & -1.052913
			& -6.161076 & -1.618110 & -4.648907 &  5.394653 &  5.997293 & -0.610818 \\
			\bottomrule
		\end{tabular}%
	}
	\\[0.5em]
	\resizebox{\textwidth}{!}{%
				\begin{tabular}{c*{12}{r}}
			\toprule
			$T~(\mu{\rm s})$
			& $a_1$ & $a_2$ & $a_3$ & $a_4$ & $a_5$ & $a_6$
			& $b_1$ & $b_2$ & $b_3$ & $b_4$ & $b_5$ & $b_6$ \\
			\midrule
			0.60
			& -4.272404 & -0.257878 & -2.750781 &  0.761435 &  1.116412 &  0.187347
			& -0.699790 &  0.540612 & -1.027274 & -2.483160 & -0.067789 &  0.669567 \\
			0.70
			& -4.594478 & -0.384034 & -3.139899 &  0.889628 &  1.209487 &  0.390048
			& -0.976911 &  0.689999 & -0.813433 & -2.280804 &  0.058660 &  0.478638 \\
			0.80
			& -4.319620 & -0.460498 & -2.865799 &  0.875479 &  0.856311 &  0.419489
			& -0.943799 &  0.679953 & -1.042947 & -2.369250 &  0.115863 &  0.407742 \\
			1.00
			& -4.248281 & -0.489297 & -3.012873 &  0.858625 &  0.689688 &  0.481482
			& -1.004111 &  0.868316 & -0.946642 & -2.243263 &  0.298252 &  0.227596 \\
			1.20
			& -3.823077 & -0.660905 & -3.089450 &  0.664824 &  0.554296 &  0.733248
			& -1.104026 &  1.132340 & -0.827626 & -2.207867 &  0.151625 &  0.041041 \\
			\bottomrule
			\end{tabular}
		}
	\caption{Optimized coefficients for Path A ramps leading to point P1 (first table) and P3 (second table)}
	\label{tab:path-coefficients}
\end{table}

\begin{figure}[ht]
	\centering
	\includegraphics[width=0.55\columnwidth]{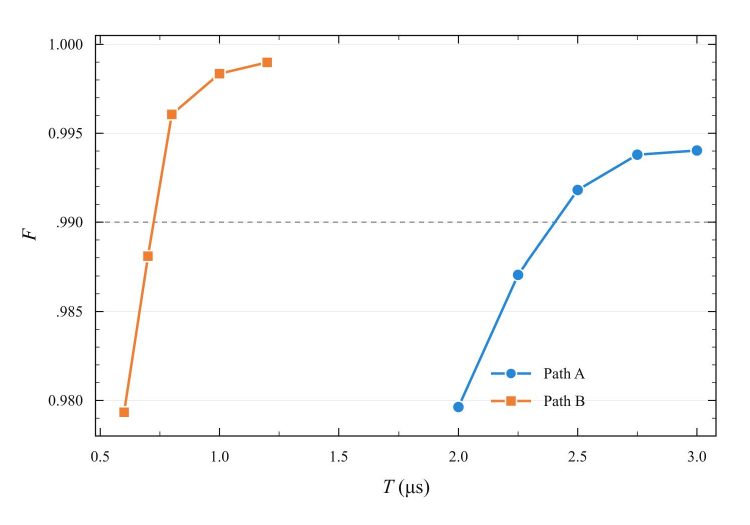}
	\caption{\label{fig:fidelity}The fidelity $F$ of the optimized ramp profiles, computed via CF4 time evolution as in App.~\ref{sec:validation}. The dashed line shows the fidelity of the ramp from \cite{Fang:2024uyf}, with $T=7\mu{\rm s}$. } 
\end{figure}

We solve this maximization problem using the Nelder-Mead method~{\cite{NelderMead1965}}
with multiple initial seeds, and settle for the best found result (which is
not necessarily a global maximum). After the optimization, we use validation from App.~\ref{sec:validation}, i.e.~the
CF4 time evolution of the full Hamiltonian to compute the true
fidelity $F$; the results are summarized in Figure~\ref{fig:fidelity}. We do the
optimization at different total times $T$. The coefficients of the optimized
ramp profiles are summarized in Table~\ref{tab:path-coefficients}. From
them, it is easy to reconstruct the ramp profile
$(\Omega' (t), \Delta' (t))$ using \eqref{eq:density}-\eqref{rho0B}. As an
example, two optimized ramp profiles (one for each path)
are shown in Fig.~\ref{fig:rampingprofile}. 

In Table~\ref{tab:fidel} we report more detailed information for two optimized Path A ramps and two optimized Path B ramps, choosing in each case the shortest and the third-shortest ramps.

\section{Deviation in 2pt function due to imperfect fidelity: theory}\label{sec:dev2ptth}

In this appendix we discuss theory for leading corrections to the spin-spin
correlation functions due to the state deviating from the exact ground state.
In Section \ref{sec:fidelity} we saw that at the end of the quasi-adiabatic evolution the
state is not the exact ground state. In Table \ref{tab:fidel} we reported, in addition
to the fidelity $| c_0 |^2$, the absolute values squared of the coefficients
$c_1$ and $c_2$ in the expansion \eqref{eq:psiexp12}. In radial quantization of CFT, the
corresponding states $| 1 \rangle $ and $| 2 \rangle $ are associated with
the states obtained by inserting CFT operators $\varepsilon$ and $\partial^2
\varepsilon$ at the origin. Specifically we have
\begin{equation}
	| 1 \rangle  = \varepsilon (0) | 0 \rangle , \quad | 2 \rangle  = \frac{1}{4
		\Delta_{\varepsilon}} \partial^2 \varepsilon (0) | 0 \rangle ,
\end{equation}
where the prefactor in the latter equation arises from the unit
norm requirement. Using the infinite-plane expressions:
\begin{equation}
	\langle \sigma (x_1) \sigma (x_2) \rangle = \frac{1}{| x_{12} |^{2
			\Delta_{\sigma}}}, \quad \langle \sigma (x_1) \sigma (x_2) \varepsilon (0)
	\rangle = \frac{C_{\sigma \sigma \varepsilon}}{| x_{12} |^{2 \Delta_{\sigma}
			- \Delta_{\varepsilon}}} \frac{1}{| x_1 |^{\Delta_{\varepsilon}} | x_2
		|^{\Delta_{\varepsilon}}},
\end{equation}
we find, for two points $x_1, x_2$ distance 1 from the origin,
\begin{align}
	\frac{\langle \sigma (x_1) \sigma (x_2) \varepsilon (0) \rangle}{\langle
		\sigma (x_1) \sigma (x_2) \rangle} &= C_{\sigma \sigma \varepsilon} | x_{12}
	|^{\Delta_e}, \\
	\frac{\langle \sigma (x_1) \sigma (x_2) \partial^2 \varepsilon (0)
		\rangle}{\langle \sigma (x_1) \sigma (x_2) \rangle} &= C_{\sigma \sigma
		\varepsilon} \Delta_{\varepsilon}^2 | x_{12} |^{\Delta_e} (4 - | x_{12} |^2)\,.
\end{align}
Substituting $| x_{12} | = 2 s, s = \sin \frac{| \phi_{12} |}{2}$, and
using $\Delta_{\varepsilon} = 1$, $C_{\sigma \sigma \varepsilon} = 1 / 2$, the
above equations give
\begin{equation}
	\frac{\langle 0 |  \sigma (x_1) \sigma (x_2) |  1 \rangle}{\langle 0 | 
		\sigma (x_1) \sigma (x_2) |  0 \rangle} = s, \quad \frac{\langle 0 |  \sigma
		(x_1) \sigma (x_2) |  2 \rangle}{\langle 0 |  \sigma (x_1) \sigma (x_2) |  0
		\rangle} = s (1 - s^2) .
\end{equation}
We can use these equations to compute relative corrections to 2pt functions for Rydberg
chain correlation functions, with $s = \sin \frac{\pi | i - j |}{N}$. The
message is that the corrections scale linearly in $c_1, c_2,
\ldots$, with order 1 coefficients. This is corroborated by the numerical
plot in Fig.~\ref{fig:rampingerror}.
	
\bibliographystyle{utphys}
\small
\setlength{\bibsep}{2pt}
\bibliography{reference.bib}

\end{document}